\documentclass{aa}

\usepackage{graphicx}
\usepackage{xcolor}
\usepackage{txfonts}
\usepackage{epstopdf}
\usepackage{hyperref}
\usepackage[font=footnotesize]{caption}
\usepackage[figure]{hypcap}

\newcommand{\Fig}[1]{Fig.~\ref{#1} }

\definecolor{ok}{HTML}{17ab3e}
\definecolor{ref}{HTML}{b5b5b5}

\institute{Leibniz-Institut f\"ur Astrophysik Potsdam (AIP), An der Sternwarte 16, D-14482 Potsdam, Germany}
\author{Y. Fournier \and R. Arlt \and D. Elstner}

\date{Accepted \dots{}. Received \dots{}; in original form \dots{}}

\title{Delayed Babcock-Leighton dynamos in the diffusion-dominated regime}

\hypersetup{
 pdfauthor={Y. Fournier, R. Arlt, D. Elstner},
 pdftitle={Delayed Babcock-Leighton dynamos in the diffusion-dominated regime},
 pdfkeywords={KEYWORD},
 pdfsubject={},
 pdfcreator={Emacs 24.5.1 (Org mode 9.0.2)}, 
 pdflang={English}}

\begin{document}

\abstract
   %
    {
      Solar dynamo models of Babcock-Leighton type typically assume the
      rise of magnetic flux tubes to be instantaneous.
      Solutions with high-magnetic-diffusivity have too short periods and
      a wrong migration of their active belts. Only the low-diffusivity
      regime with advective meridional flows is usually considered.
    }
   %
    {
      In the present paper we discuss these assumptions and applied
      a time delay in the source term of the azimuthally averaged induction equation.
      This delay is set to be the rise time of magnetic flux tubes which
      supposedly form at the tachocline.
      We study the effect of the delay, which
      adds to the spacial non-locality a non-linear temporal one,
      in the advective but particularly
      in the diffusive regime.
    }
   %
    {
      \cite{Fournier+17} obtained the rise time according to stellar
      parameters such as rotation, and the magnetic field strength
      at the bottom of the convection zone. These results allowed us to
      constrain the delay in the mean-field model used in a parameter
      study.
    }
   %
    {
      We identify
      an unknown family of solutions. These solutions self-quench,
      and exhibit longer periods than their non-delayed counterparts.
      Additionally, we demonstrate that the non-linear delay is responsible
      for the recover of the equatorward migration of the active belts
      at high turbulent diffusivities.
    }
   %
    {
      By introducing a non-linear temporal non-locality (the delay)
      in a Babcock-Leighton dynamo model, we could obtain solutions
      quantitatively comparable to the solar butterfly diagram
      in the diffusion-dominated regime.
    }


\maketitle

\section{Introduction}
\label{sec:introduction}

The magnetic solar cycle is attributed to a dynamo process
in which motions of a conductive medium are leading to the
continuous induction of magnetic fields. Those motions may
arise from convection which becomes anisotropic in the
presence of rotation and stratification and is approximatively
described by the $\alpha$-effect \citep{KrauseRaedler80}, or
from the rise of magnetic structures to the solar surface,
again in the presence of rotation and stratification, leading
to what is called the Babcock-Leighton effect \citep{Leighton69}.
While there has been no derivation of this effect from first 
principles so far, the synergy with the meridional circulation 
in the convection zone may lead to a periodic magnetic field
explaining the solar cycle satisfactorily. 

Babcock-Leighton dynamo were shown to reproduce qualitatively 
the solar butterfly diagram only if the turbulent magnetic 
diffusivity in the bulk of the convection zone is less than
$10^{12}$~cm$^2$/s, a regime often referred to as the advective
regime, since the meridional circulation then determines the
cycle period. In the diffusive regime, the cycle period varies
with the turbulent magnetic diffusivity and can be reduced to 
the observed cycle length, but the propagation of the dynamo
wave then follows the Parker-Yoshimura rule and is poleward
at all latitudes \citep{Yoshimura75}.

A variety of Babcock-Leighton dynamos has been published
during the last 25~years,
exemplarily by \citet{Choudhuri+95}, \citet{Dikpati+99}, 
\citet{Kueker+01}, \citet{Chatterjee+04}, \citet{Guerrero+07}, and \citet{Sanchez+14} 
for kinematic models with stationary flows and dynamo effect, 
\citet{Nandy+01} for a model with toroidal-field 
loss by buoyancy, as well as \citet{Kitchatinov+11}
for a nonlocal $\alpha$-effect similar to the 
Babcock-Leighton effect. A non-kinematic simulation showing
a solar-like butterfly diagram is the one by \citet{Rempel06}
with a turbulent magnetic diffusivity rising from $10^{10}$~cm$^2$/s
to $10^{12}$~cm$^2$/s in the convection zone. Beyond the Sun,
and among others, \citet{Jouve+10a} studied the cycle period 
dependence of Babcock-Leighton dynamos on the stellar rotation 
rate. The variability of 
the solar cycle including grand minima has been addressed with 
Babcock-Leighton dynamos by, for example, \citet{Karak10} who
varied the meridional circulation, \citet{Olemskoy+13}
who varied the strength of their nonlocal source term,
and \citet{Inceoglu+17} who varied both the generation
of the flows and the Babcock-Leighton effect in a non-kinematic
setup. In all these attempts, however, the effect of the toroidal
magnetic field in the interior of the convection zone on the
poloidal-field generation at the surface is instantaneous.
Note that this list of Babcock-Leighton type dynamo papers is far 
from complete.

In the present Paper, we are addressing the problem that the 
Babcock-Leighton effect is not only nonlocal in space, but
also in time. Following the pioneering work by \citet{Jouve+10b},
we take a further step toward a fully constrained
Babcock-Leighton dynamo. Here we use the results of global
numerical simulations of flux-tube rise to improve and actually 
constrain a Babcock-Leighton dynamo model.

The global numerical simulations by \citet{Fournier+17} have 
shown that the rise time of magnetic flux tubes is independent 
of the magnetic diffusion. We suggest to treat this independence 
with a non-locality in space and time, by designing a 
Babcock-Leighton dynamo model which has a time delay in the 
source term.

Non-localities have been shown to generate long-term variability
in the amplitude and in the period of the magnetic cycle for
a wide variety of models. A remarkable early attempt is the one
by \citet{Yoshimura78} who used already a two-dimensional setup
with source terms nonlocal both in space and time, delivering
cyclic magnetic fields interrupted by ``low-activity'' periods of
a few cycles duration. A sequence of papers on a zero-dimensional
model, i.e. without any spatial dependence but with time delay, 
was published by \citet{Wilmot-Smith+06}, \citet{Hazra+14} 
(including stochastic variations in the Babcock-Leighton effect), 
and \citet{Tripathi+18}, who found sub-critical dynamo action,
hinting on what we are going to show in this Paper in a two-dimensional
spherical shell with realistic solar differential rotation.
This is in line with the result by \citet{Rheinhardt+12} who
studied the memory effect in a turbulent-$\alpha$ dynamo
and found a threshold for dynamo excitation which is lower 
than in the case of no memory effect. There is actually a variety
of papers on non-localities in turbulent-$\alpha$ dynamos
which we do not review here. Closest to our approach is the 
work by \citet{Jouve+10b} who made the time delay magnetic-field 
dependent. However the authors always found that the time 
correlations required to obtain solar-like variability were 
too long to agree with the model's assumptions.

Here we study such non-local models with time delays depending 
on the magnetic field strength, where the dependence is 
derived from the flux-rise simulations by \cite{Fournier+17}. 
Section~\ref{sec:the-model} describes the equations used and 
Section~\ref{sec:the-delayed-dynamos} shows the general results 
of using a delay in the induction equation. We discuss the 
implications of the results in Section~\ref{sec:discussion-and-conclusions}.

\section{The model and the reference setup}
\label{sec:the-model}

We derive our model from \cite{Jouve+08}, and like in \cite{Jouve+10b}
we introduce a delay into the Babcock-Leighton source term of a mean-field dynamo.
In the current work, we will present the results of a large parameter study of more than 2000 simulations,
partly constrained thanks to the results of global numerical simulations.

The model used here is simplified: we assume a constant turbulent magnetic
diffusivity, $\eta_{\rm t}={\rm const}$, in the stellar interior. Since the
turbulent diffusivity is a measure of the turbulence intensity, a constant
value implies that we can neglect radial turbulent pumping \citep{Raedler68}.
This choice is made to prevent our setup from being polluted with weakly constrained
parameters, namely an arbitrary profile of the magnetic diffusivity.
Indeed, observations of decaying active regions suggest that
the turbulent magnetic diffusivity, $\eta_{\rm t}$, at the surface is of the order
of $10^{12} {\rm cm}^{2}/{\rm s}$, while the mixing length theory provides an upper limit
of $10^{14} {\rm cm}^{2}/{\rm s}$ at the surface and $10^{13} {\rm cm}^{2}/{\rm s}$
at the bottom of the convection zone.
The large discrepancy between the various available estimates demonstrates the limit of our current
knowledge. 
Additionally a constant turbulent magnetic diffusivity has the appreciable side effect that
the Reynolds number is small everywhere allowing coarser grids and shorter computation times,
which is adequate for a large parameter study.

Lengths and time are normalized with the stellar radius, $R_{\star}$, and the
turbulent magnetic diffusion time, $\tau_{\rm diff} = R^{2}_{\star}/\eta_{\rm t}$. 
While the free choice of the unit of the magnetic flux density $\vec B$ is made
by setting the equipartition value with convective motions $u_{\rm rms}$ at $r = 0.71$,
$B_{\rm eq} = u_{\rm rms} \sqrt{\mu_0 \rho}$ to unity, where $\rho$ and $\mu_0$
are the gas density and the permeability constant, respectively (see below).
The resulting dimensionless set of equations can be written in spherical coordinates $(r, \theta, \phi)$ as:   
%
\begin{eqnarray}
  \label{eq:system}
  \partial_t B_{\phi} ~&=&~ \left( \nabla^2 - \frac{1}{\varpi^2} \right) B_{\phi} \nonumber\\
                    &&~ - ~ {\rm Re} \, \varpi {\bold u}_{\rm P}
                            \cdot \nabla\left(\frac{B_{\phi}}{\varpi}\right)
                      ~ - ~ {\rm Re} \, (B_{\phi} \nabla) \cdot {\bold u}_{\rm P} \nonumber\\
                    &&~ + ~ C_\Omega \, \varpi \,
                           \left[\nabla \times (\varpi A_{\phi} {\bold e}_{\phi}) \right]
                           \cdot \nabla(\Omega) {\rm ,} \nonumber\\
                           \nonumber\\
  \partial_t A_{\rm \phi} ~&=&~ \left( \nabla^2 - \frac{1}{\varpi^2} \right) A_{\phi} \nonumber\\
                          &&~ - ~ {\rm Re} \, \frac{{\bold u}_{\rm P}}{\varpi} \cdot \nabla (\varpi A_{\phi})
                          ~ + ~ C_{\!S} S {\rm ,}
\end{eqnarray}
%
with $B_{\phi}$, ${\bold u}_{\rm P}$, $\varpi$, $A_{\phi}$, $\Omega$
and $S$, being the azimuthal magnetic flux density, the meridional
circulation profile, the cylindrical distance to the rotation axis,
the vector potential of the poloidal magnetic field, the angular velocity,
and the source term for the poloidal field, respectively.

This system is controlled by three dimensionless parameters, the
Reynolds numbers for the rotation and the meridional flow, $C_{\Omega}$
and ${\rm Re}$, and the dynamo number $C_{\!S}$.
%
\begin{equation}
   {\rm Re} = \frac{u_0 R_\star}{\eta_{\rm t}}            ~{\rm ; }~~
   C_\Omega = \frac{\Omega_0 R^2_{\star}}{\eta_{\rm t}} ~{\rm ; }~~
   C_{\!S} = \frac{S_{\!0} R_{\star}}{B_{\rm eq} \eta_{\rm t}}   ~{\rm , }
\end{equation}
%
with $u_0$, $\Omega_0$, $S_{\!0}$ being the maximum
meridional velocity, angular velocity and Babcock-Leighton 
effect, respectively. The profile of ${\bold u}_{\rm P}$ and $\Omega$
are normalized by $u_0$ and $\Omega_0$, respectively.

\begin{figure}[tbph!]
\centering
  \begin{tabular}{l}
    \includegraphics[width=.9
      \linewidth]{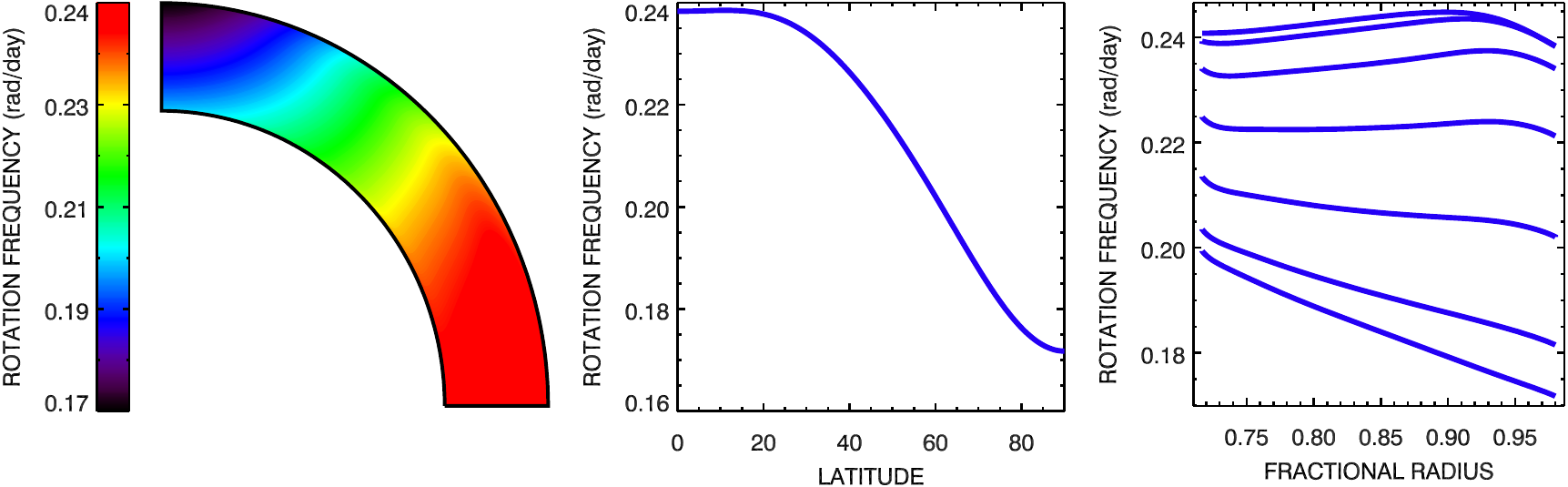} \\
    \includegraphics[width=.6\linewidth]{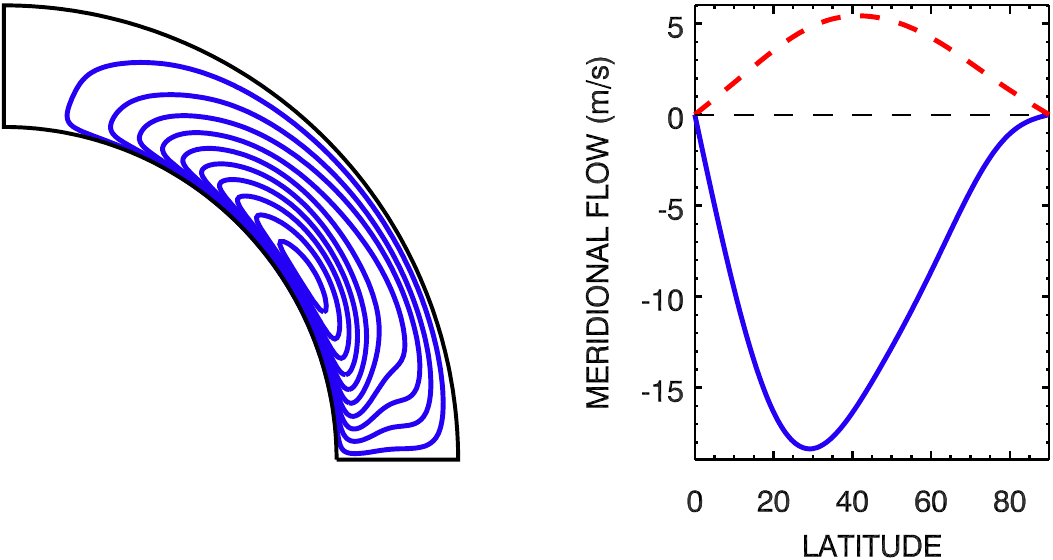} \\
  \end{tabular}
  \caption[Solar-like differential rotation and meridional circulation,
           obtain from a mean-field simulation of the $\Lambda$-effect.]{
           \label{fig:diffrot-meridional}
           Solar-like differential rotation and meridional circulation,
           obtained from a mean-field hydrodynamic simulation \citep{Kuker+11}.
          }
\end{figure}

The loose constraint on the amplitude and the profile of the
meridional circulation gives some freedom for the choice of ${\rm Re}$.
The original model used a canonical profile aiming to catch the general
characteristics of the solar flows.
However the profile remained arbitrary, and the surface shear layer was missing.
We suggest to take advantage of the results of \cite{Kuker+11},
where the authors provide consistent profiles of differential rotation and meridional flows based on the
$\Lambda$-effect theory \citep{Rudiger89}. This theoretical result provides a solar-like profile
including a surface shear layer, with a single free coefficient,
whose value is set to fit the helioseismic observations. The differential rotation and
meridional circulation profiles and amplitudes are illustrated in Fig.~\ref{fig:diffrot-meridional}.
%

In Babcock-Leighton dynamos, the source term $S$ is based on phenomenological arguments.
Observations suggest that the reversed polarity of the poloidal field emerging
at the surface leads to a reversal of the large scale polar magnetic field.
The source term $S$ is an attempt to catch the physical processes behind the generation of
poloidal field from the deeply seated toroidal field.

In the traditional frame of the Babcock-Leighton effect, for sufficiently
large magnetic fields, buoyancy transports toroidal magnetic flux to the surface.
During its rise through the convection zone,
poloidal field can be generated under the action of the Coriolis effect, providing the necessary field at the surface
for the Babcock-Leighton effect to work.
Global simulations could show that buoyant magnetic structures
locally quench the magnetic turbulent diffusivity, $\eta_{\rm t}$,
allowing them to remain coherent along their rise \citep{Cattaneo+88}.
Additionally, \cite{Fan+94} showed that the solar meridional
flow does not significantly affect the rise of magnetic flux tubes.
Therefore, the buoyant transport of magnetic flux is independent of
$\eta_{\rm t}$ and of the meridional flow, ${\bold u}_{\rm P}$,
and depends exclusively on the magnetic pressure, i.e. on $B$.

In the model presented here, we treat the transport's independence of
$\eta_{\rm t}$ as a memory effect in
the mean-field equations.
The source of the poloidal field at the surface is correlated with the toroidal
field at the bottom of the convection zone at an earlier time.
The considered time correlation, $\tau_{\rm delay}$, represents the time required
by a buoyant magnetic structure to rise from the bottom of the convection zone
to the surface.

We write the delayed source term as follow:
%
\begin{eqnarray}
  \label{eq:source-term}
  S(r, \theta, t)
  &=& f(r,\theta)
\sum_{\tau_{i}} \left\{
                \left[
    1 + \left(
                  B_{\phi}(r_0, \theta, t - \tau_{i})\,/\,B_{\rm quench}
    \right)^2
                \right]^{-1}\right.\nonumber\\
  & &\phantom{f(r,\theta) \sum_{\tau_{i}}}
     B_{\phi}(r_0, \theta, t - \tau_{i})
     \Bigg\}{\rm ,} \nonumber\\
\nonumber\\
{\rm for} &~& B_{\phi}(r_0, \theta, t - \tau_{i}) > B_{\rm threshold} {\rm ,}~
S=0~{\rm otherwise.}
\end{eqnarray}
%
Where the sum is computed over all magnetic flux tubes reaching the surface
at a given time $t$.

On the one hand, tubes with magnetic flux densities larger than $B_{\rm quench}$
are weakly affected by the Coriolis effect and emerge as untilted active regions
\citep{DSilva+93, Jouve+13, Fournier+17}.
Such untilted regions do not provide poloidal magnetic flux to the Babcock-Leighton
effect.
On the other hand, weak flux tubes may be strongly affected by the turbulent
convective motions such that they won't reach the surface. We consider a threshold,
$B_{\rm threshold}$, below which the toroidal field does not participate in the
dynamo mechanism. Such a lower limit prevents the dynamo from growing
from an arbitrarily low seed field. 

Since the place where the source $S$ operates remains unknown,
we use the same arbitrary profile of \cite{Jouve+08}, where $S$
is nonzero at $r\goa 0.9$ and is maximum at $45^{\circ}$ latitude at the surface.
%
\begin{equation}
  f(r, \theta) = \frac{1}{2} ~ \left[ 1
                         + {\rm erf} \left( \frac{ r - 0.9 }{0.02}
                           \right)
                               \right]
                         \cos\,\theta \sin\,\theta ~{\rm .}
\end{equation}

\subsection{Introducing the 3D results into the model}
\label{sec:modelling-the-delay}

Thanks to global simulations it is now possible to constrain
the correlation time, $\tau_{\rm delay}$, the source-term quenching, $B_{\rm quench}$,
and the magnetic field threshold, $B_{\rm threshold}$.

Here $\tau_{\rm delay}$ represents the rise time of magnetic flux tubes, 
i.e. coherent magnetic structures which may form from the destabilisation
of a previously amplified magnetic layer \citep{Rempel+01, Hotta+12}, by the
buoyancy instability \citep{Parker55, Matthews+95, Wissink+00, Fan01, Kersale+07, Favier+12}.
The amplification depends only on the stratification of the solar convection
zone, and has been found to be $F_{\rm amp} = 10$ \citep{Rempel+01, Hotta+12}.
As a result amplified flux tubes can reach up to $15\cdot10^{4} \, {\rm G}$ (i.e. $10 \, B_{\rm eq}$).

\cite{Fournier+17} demonstrated that the rise time of magnetic
flux tubes follows the relation:
\begin{equation}
  {\tau}_{\rm delay} \propto P_{\rm rot} \left( \Gamma_{\alpha_1}^{\alpha_2} \right)^{\alpha_3} ~~{\rm ,}
\end{equation}
where $P_{\rm rot}$ is the rotation period of the star, and $\Gamma_{\alpha_1}^{\alpha_2}$
is the ratio between the buoyant force and the Coriolis effect, modified
by the magnetic tension. 
The exponent $\alpha_3$ is a function of the azimuthal mode number with which the magnetic flux tube rises.
In the case of the Sun, taking $P_{\rm rot}$ constant, this relation can reduces to:
\begin{equation}
  \label{eq:delay-model}
  {\tau}_{\rm delay} \propto \tau_{0} \left| \frac{B_{\phi}}{B_{\rm eq}} \right|^{\alpha} ~~{\rm ,}
\end{equation}
with $\alpha$ varying between $-0.91$ and $-2.0$ depending on the azimuthal mode.
The parameter $\tau_0$ is the rise time for a field in equipartition with the 
convective velocity, $B_{\phi} = B_{\rm eq}$ and is a free parameter. It depends 
on the rotation period of the star, the depth of its convection zone and
the details of the destabilisation process forming the flux tubes
as well as on the profiles of the turbulent thermal conductivity, viscosity 
and diffusivity. A discussion can be found in the Appendix.

In Eq.~\ref{eq:delay-model} there is no latitudinal dependence. But recalling that this equation
describes the reduction of the rise velocity by the tension force
we can model it following the latitudinal dependence of the
tension force of $1 / \sin \theta$.
\begin{equation}
  {\tau}_{\rm delay} = \frac{\tau_{0}}{\sin \theta} \left| \frac{B_{\phi}}{B_{\rm eq}} \right|^{\, \alpha} ~ {\rm .}
\end{equation}

We further define an effective delay ${\tau}_{\rm eff}$, which turns out to be a very useful parameter.
It corresponds to the shortest delay in any given moment, i.e.
\begin{equation}
  {\tau}_{\rm eff} = \tau_{0} \left| \frac{B^{\rm max}_{\phi}}{B_{\rm eq}} \right|^{\, \alpha} ~ {\rm ,}
\end{equation}
where $B^{\rm max}_{\phi}$ is the strongest field strength generated by the $\Omega$-effect
at the bottom of the convection zone. The maximum is taken over time and latitude (after saturation).

\cite{Rempel06} showed that Babcock-Leighton dynamos provide
$B_{\phi}$ up to $3 B_{\rm eq}$.
Since the resulting amplified flux tubes are in the buoyancy dominated regime and therefore weakly
participate in the dynamo because the Coriolis effect is not strong enough for significantly tilted
active regions (see Appendix), we set $B_{\rm quench} = 3 \, B_{\rm eq}$.
\cite{Fan+03} have shown that amplified flux tubes weaker than $3 B_{\rm eq}$ will not
reach the surface as coherent structures. The lower threshold on $B_{\phi}$
is $B_{\rm threshold} = 3 B_{\rm eq} / F_{\rm amp} = 0.3 B_{\rm eq}$. Since magnetic flux tubes
of threshold flux density may have rise times of the order of a cycle, we 
implemented a source term ``buffer'' lasting until $t + 5 \tau_{\rm diff}$ for any
moment $t$ in the simulation, in order to prevent any issues with very long delays.

\subsection{Behavior of the delay}
\label{sec:behavior-of-the-delay}

\begin{figure*}[tbph!]
\centering
  \begin{tabular}{cc}
    \includegraphics[width=0.48\linewidth]{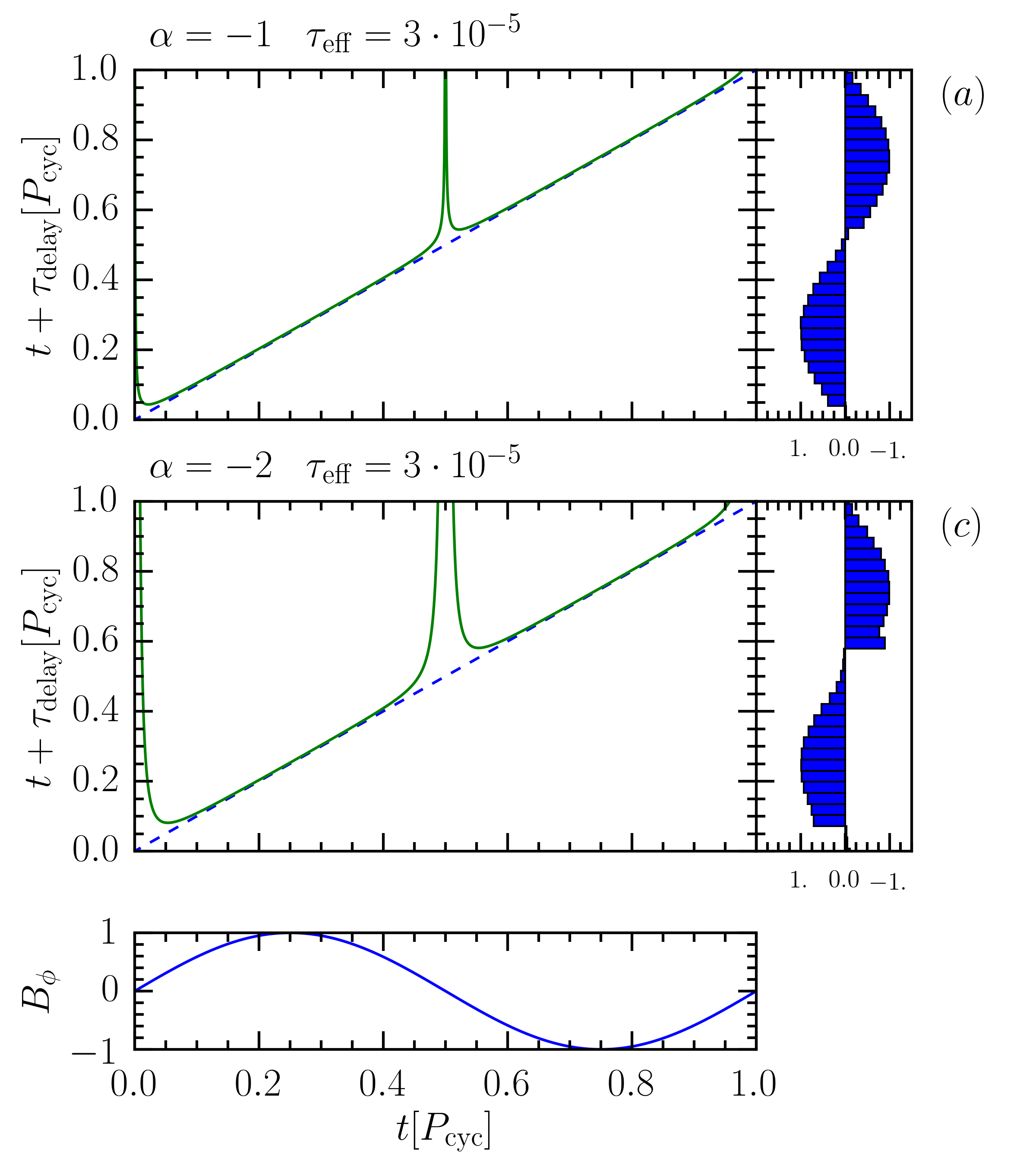} &
    \includegraphics[width=0.48\linewidth]{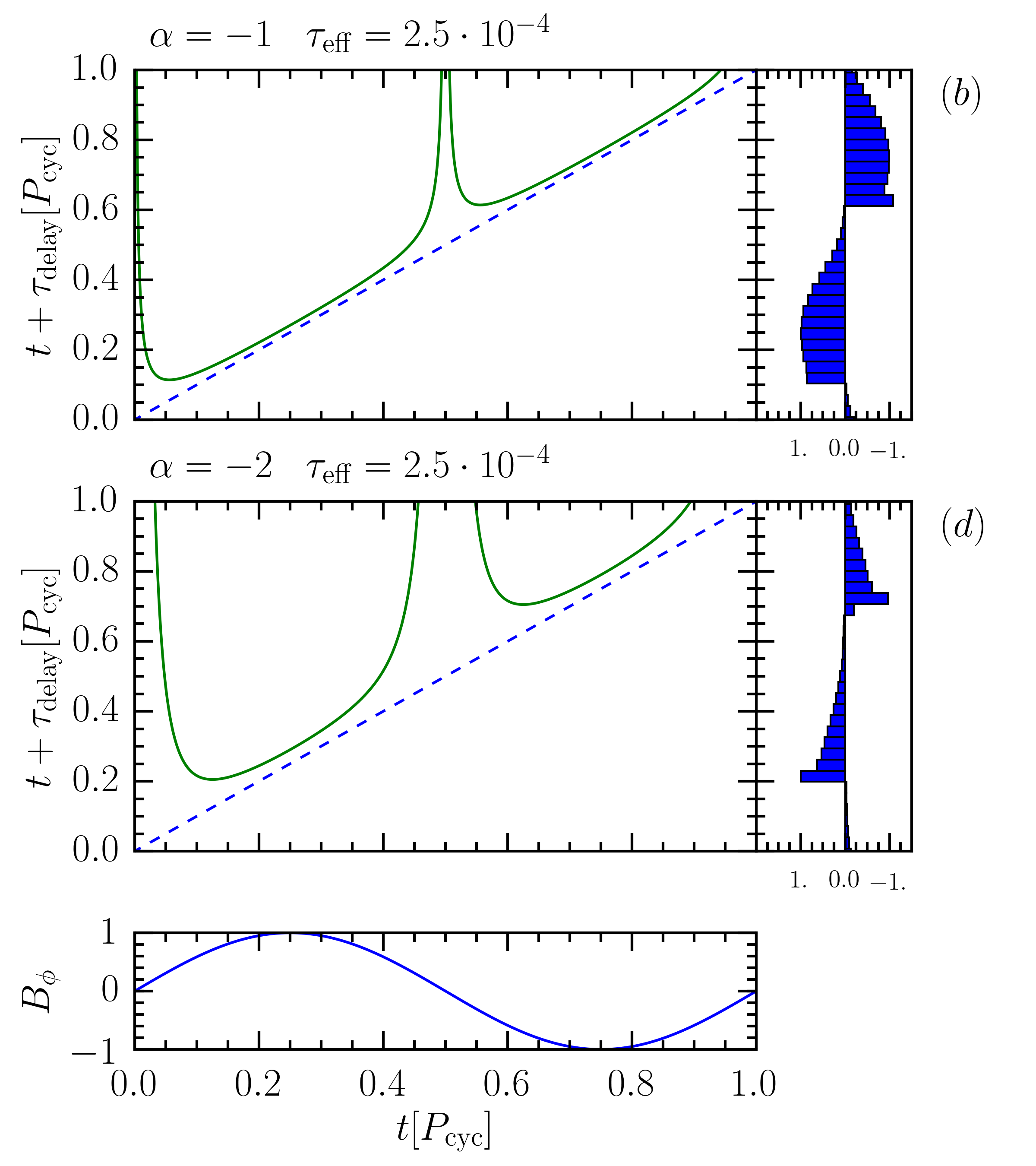} \\    
  \end{tabular}
  \caption[]{
           \label{fig:theoretical-delay}
           Zero-dimensional toy model illustrating the effect of the delay.
           On each panel the green curve represents the evolution of the delay
           with time for a given $B_{\phi}$, indicated in the lower panel as the blue curve.
           The dashed line represents time. The histogram on the
           right-hand side represents the resulting source term $S$ normalized in arbitrary unit.
           Panels (a) and (b), illustrate the case of a weakly non-linear delay ($\alpha = -1$),
           whereas (c) and (d) a non-linear delay ($\alpha = -2$).
           Panels (a) and (c) with a short effective delay (e.g.: $B^{\rm max}_{\phi} = B_{\rm eq}$ and $\tau_0 = 3 \cdot 10^{-5} P_{\rm cyc}$)
           and panels (b) and (d) with a long effective delay (e.g. $B^{\rm max}_{\phi} = 0.5 \, B_{\rm eq}$ and $\tau_0 = 10^{-3} P_{\rm cyc}$).
          }
\end{figure*}
%
\begin{figure}[tbph!]
  \centering
  \begin{tabular}{l}
    \includegraphics[width=1.0\linewidth]{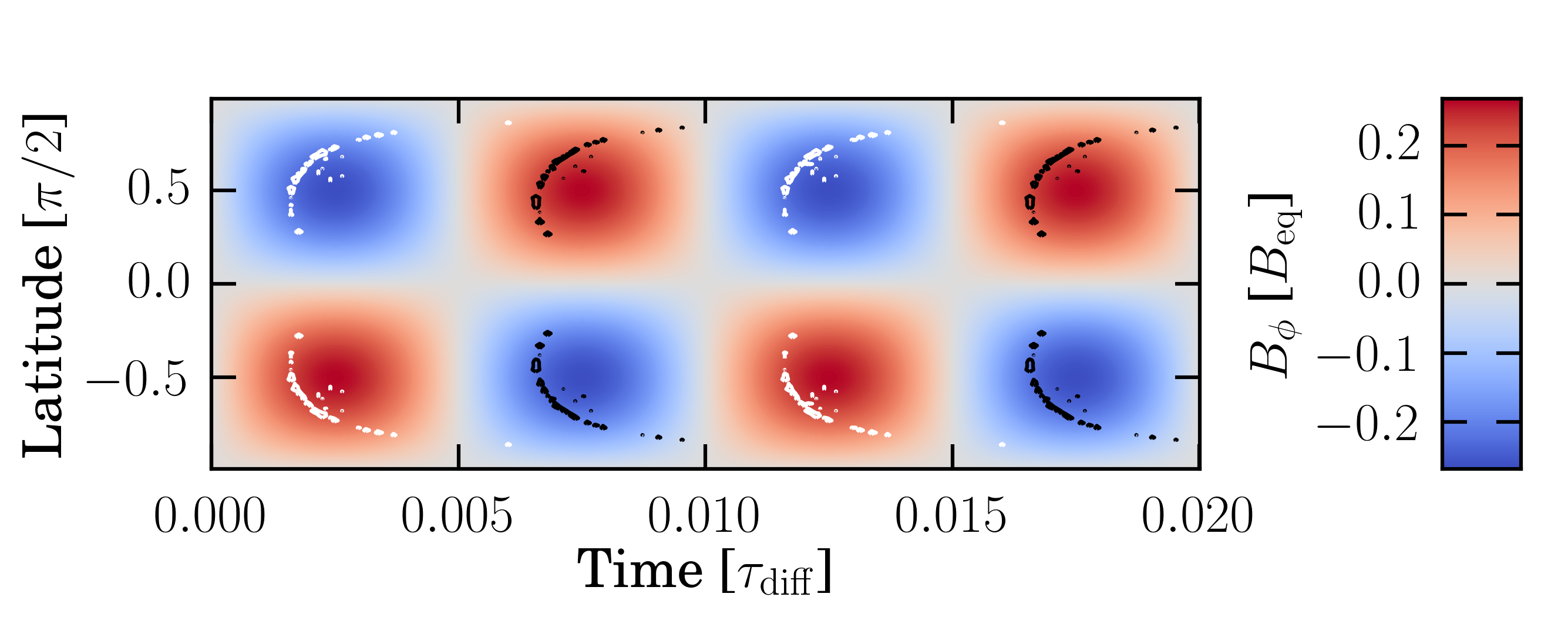} \\
    \includegraphics[width=1.0\linewidth]{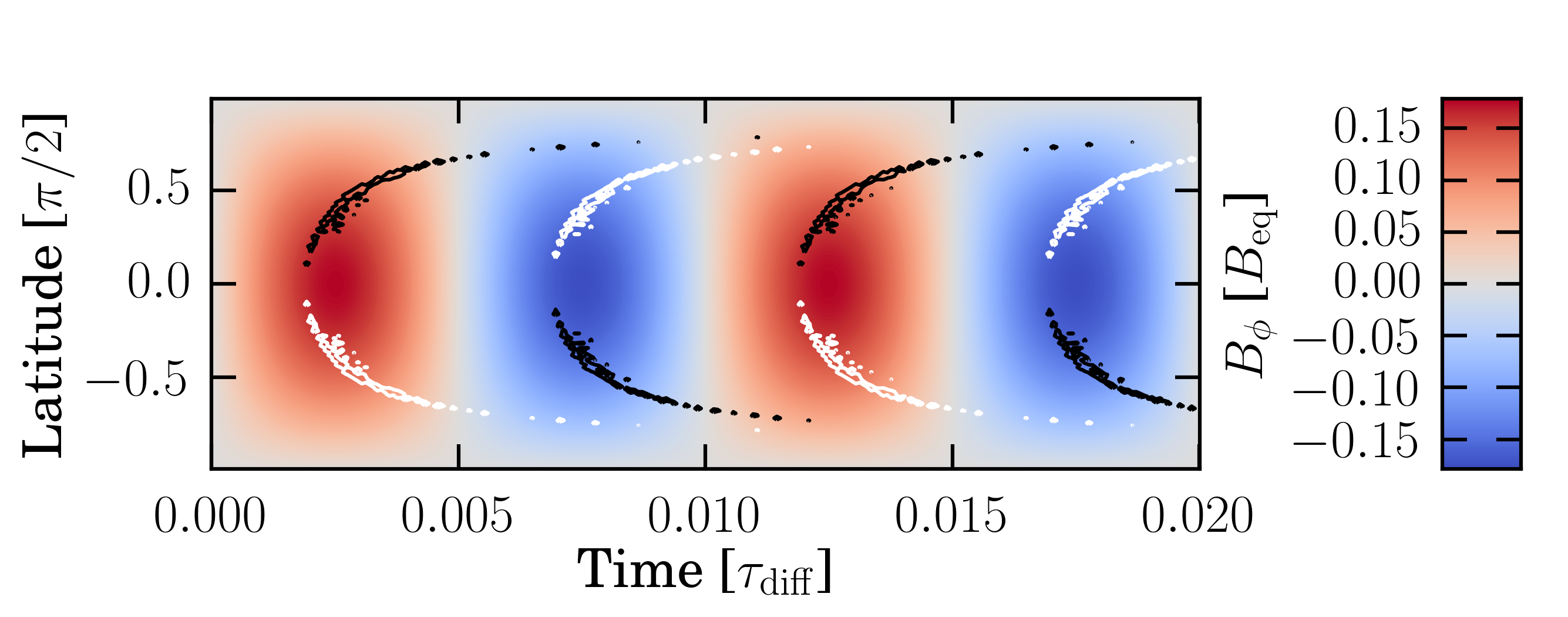} \\
  \end{tabular}
  \caption[]{
           \label{fig:theoretical-latitudinal-profile-of-the-delay}
           Colour coding of the toroidal magnetic field strengths of an artificial 
           oscillatory dipole (top) and an artificial oscillatory quadrupole (bottom).
           The contours represent the source term ($S$) at the
           surface for $\alpha = -2$ and $\tau_0 = 10^{-2} P_{\rm cyc}$. 
          }
\end{figure}
%
The dependence of the delay on $B_{\phi}$ is controlled by $\alpha$.
When $\alpha$ is set to zero, the delay is constant.
A constantly delayed source term has the same time-profile
as a non-delayed source term, but is shifted in time.
However as soon as $\alpha$ becomes non-zero
the delay becomes time-dependent. Weaker flux tubes rise longer
than stronger ones. Weak flux tubes may therefore reach the surface at the
same time than stronger flux tubes formed at a later time.
The time-dependence of the delay results into an accumulation of
the source at the surface around certain times.

We illustrate this behavior in Fig.~\ref{fig:theoretical-delay},
for four different idealized cases. We consider a single point with
a sinusoidally varying toroidal magnetic field,
\begin{equation}
  B_{\phi} = B^{\rm max}_{\phi} \sin \left(\omega \, t \right) ~~ {\rm ,}
\end{equation}
with amplitude $B^{\rm max}_{\phi}$ and frequency $\omega$. 
The field is shown as blue curves in Fig.~\ref{fig:theoretical-delay}, 
while the red curves are the delay computed from $B_{\phi}$.
The histograms show the resulting source of poloidal field 
accumulated from $B_{\phi}$ in the past.
Both upper panels show the case of a weakly non-linear, short- and a long-$\tau_{\rm eff}$, respectively
(with $\alpha = -1$).
The two lower panels illustrate a strongly non-linear short- and long-$\tau_{\rm eff}$
(with $\alpha = -2$).
In the weakly non-linear, short-$\tau_{\rm eff}$ case the accumulation seen in the diagram
is almost negligible and the time-profile of the source looks almost identical
to the profile of $B_{\phi}$.
It is only for a longer $\tau_{\rm eff}$ that the accumulation becomes visible.
A similar accumulation is found for the non-linear, short-$\tau_{\rm eff}$,
demonstrating that the non-linearity support the accumulation.
In the case of the long-$\tau_{\rm eff}$ the deformation of the profile
leads to a ``front''.

Clearly, there are two aspects of the delay which lead to accumulation, the length of the
effective delay, $\tau_{\rm eff}$, controlled by $\tau_0$ and $B^{\rm max}_{\phi}$,
and its non-linearity, controlled by $\alpha$.

Because of the latitudinal dependence of the toroidal magnetic field,
the delay naturally varies with latitude. Like its time dependence, the latitudinal
dependence of the delay is a function of $\alpha$ and $\tau_0$, but it additionally depends
on the sign of the latitudinal gradient of the toroidal field, $\partial B_{\phi} / \partial \theta$.
As it can be seen in Fig.~\ref{fig:theoretical-latitudinal-profile-of-the-delay},
if the toroidal field decreases toward the equator, the resulting delayed field
-- illustrated by the contour plots -- migrates equatorward, whereas when the gradient
is directed poleward the dynamo wave propagates poleward. 
This is due to the fact that weaker fields are rising on a longer
time scale and the accumulation is retarded setting the direction of
propagation with the latitudinal gradient of $B_{\phi}$.

\section{The non-delayed model}

\begin{table*}[tbh]
  \caption[]{\label{tab:setups} parameters of the various setups.}
  \centering
  \begin{tabular}{rccr|rrccccc}
    \hline \hline
    Setup  & $\Omega_0$          & $u_{0}$ &\multicolumn{1}{c|}{$\eta_{\rm t}$}    &\multicolumn{1}{c}{$C_{\Omega}$}&\multicolumn{1}{c}{${\rm Re}$}& 
    $C_{S}$           & $\tau_0$            & $B_{\rm quench}$ & $B_{\rm threshold}$ & $F_{\rm amp}$ \\
           & [s$^{-1}$]          & [m/s]   &\multicolumn{1}{c|}{[cm$^2$/s]}       &                &            &                   & $[\tau_{\rm diff}]$ & $[B_{\rm eq}]$   & $[B_{\rm eq}]$&\\
    \hline 
    ADV    & $2.68\cdot 10^{-6}$ & $14.37$ & $10^{11}$         & $1.3\cdot10^5$ & $1000$     & \phantom{1}$3.73$ & $0.$                & $32$    & 0.1    & 10. \\
    DIFF   & $2.68\cdot 10^{-6}$ & $14.37$ & $10^{12}$         & $1.3\cdot10^4$ &  $100$     & $12.5$\phantom{1} & $0.$                & $32$    & 0.1    & 10. \\
    \hline
    D-ADV  & $2.68\cdot 10^{-6}$ & $14.37$ & $10^{11}$         & $1.3\cdot10^5$ & $1000$     & varied            & varied              & $32$    & 0.1    & 10. \\
    D-DIFF & $2.68\cdot 10^{-6}$ & $14.37$ & $10^{12}$         & $1.3\cdot10^4$ &  $100$     & varied            & varied              & $32$    & 0.1    & 10. \\
    \hline
    SOLAR  & $2.68\cdot 10^{-6}$ & $14.45$ & $6.7\cdot10^{11}$ & $1.95\cdot10^4$ & $150$     &\phantom{1}$5.18$  & $10^{-3}$           & $32$    & 0.1    & 10. \\
    \hline
  \end{tabular}
\end{table*}

\begin{figure}[tbph!]
  \centering
  \begin{tabular}{l}
    \includegraphics[width=1.0\linewidth]{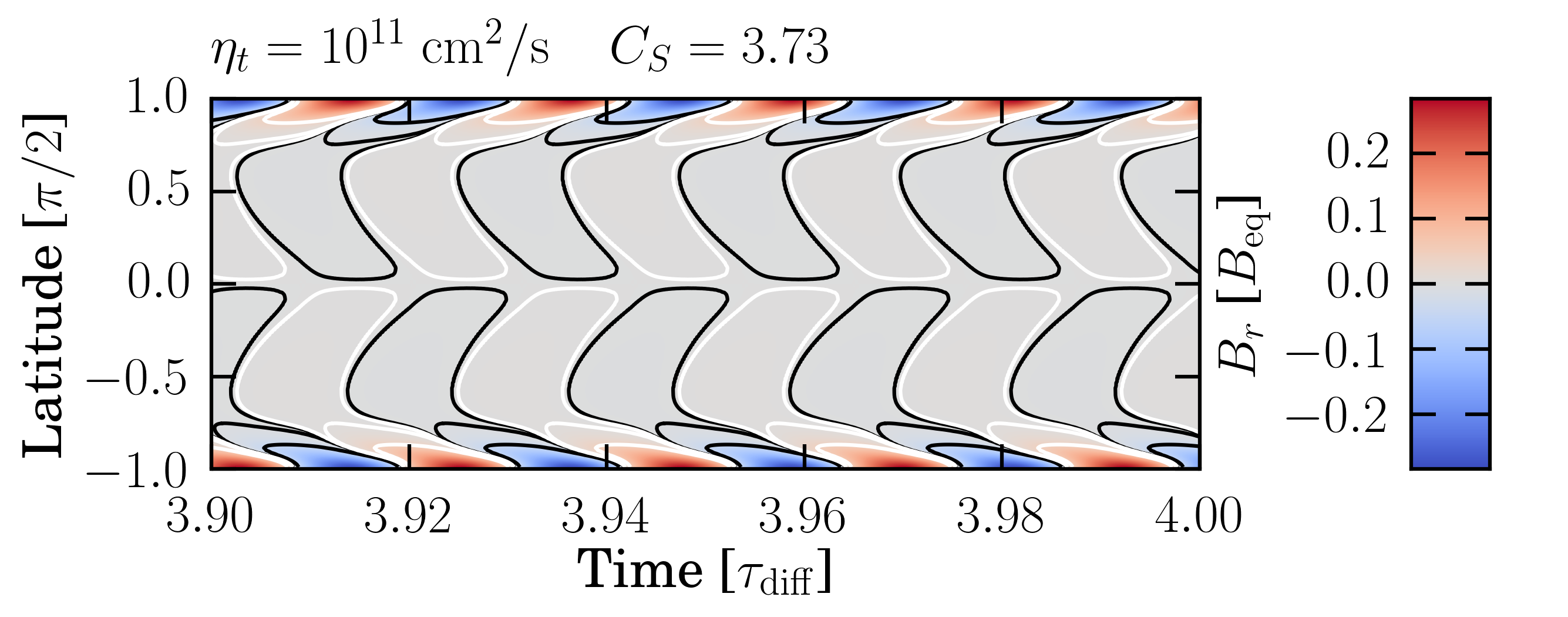} \\
    \includegraphics[width=1.0\linewidth]{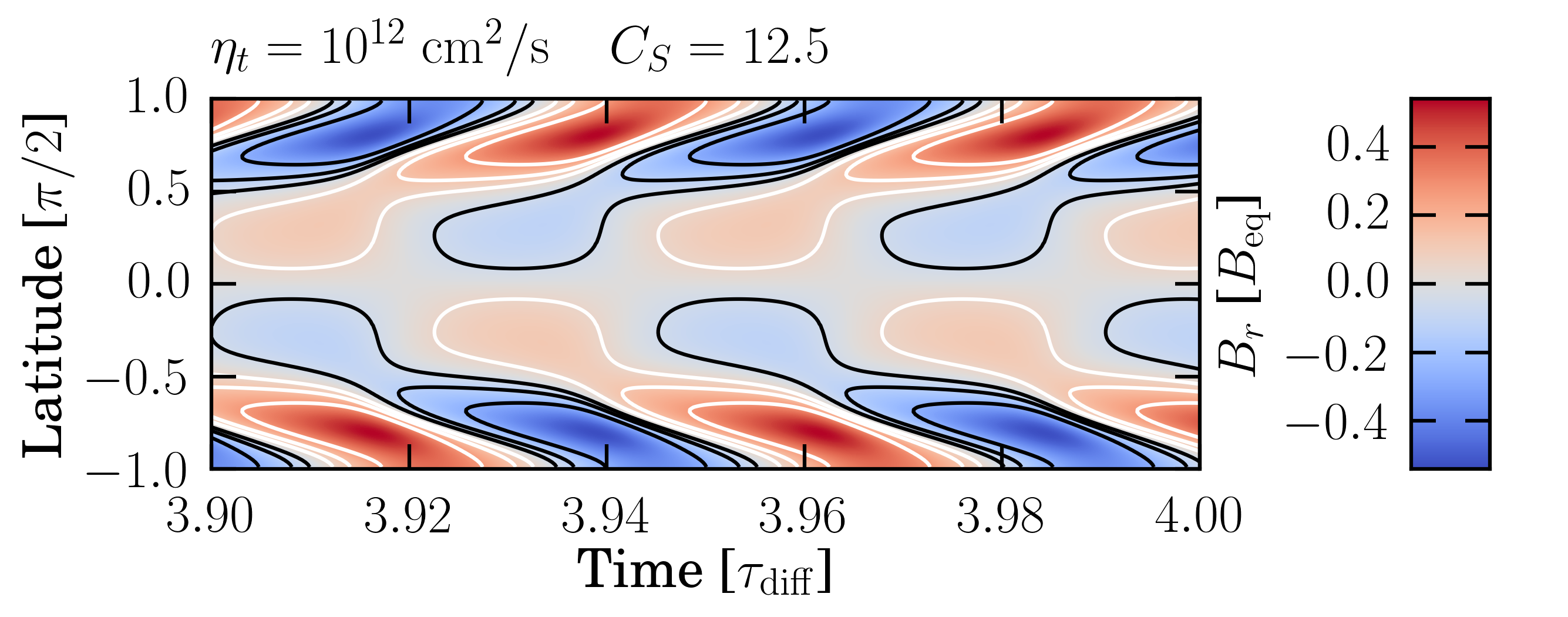} \\
  \end{tabular}
\caption[]{
           \label{fig:non-delayed-model}
           Radial magnetic field at the surface as a function of time from the solutions in the advection dominated
           regime (ADV, top) and the diffusion dominated regime (DIFF, bottom).
           Contours also represent $B_r$, but show the 15, 1.5 and 0.15\% levels of $B_{r}^{\rm max}$ 
           for the top panel (ADV) and the 50, 25 and 10\% levels of $B_{r}^{\rm max}$ for
           the bottom panel (DIFF).
           The parameters are summarized in Tab.~\ref{tab:setups}.
          }
\end{figure}

We solve Eqs.~\ref{eq:system} with the pseudo-spectral, spherical code 
by \citet{Hollerbach00}, in which the diffusion term is solved in spectral space, while the
induction term is solved in real space. We do not
employ the momentum and temperature equations of the code here.
The induction equation solution took part in the benchmark by 
\cite{Jouve+08}, and after implementing the delay term for the present
study, we confirmed their results for $\tau_{\rm delay}=0$
as is shown in Fig.~\ref{fig:non-delayed-model}.
Our
constant turbulent diffusivity of $\eta_{\rm t} = 10^{11} {\rm cm}^{2} / {\rm s}$
as compared to the radius-dependent one in the benchmark,
and our theory-based differential rotation versus the closed approximation in the benchmark do not modify the overall picture of the
solutions.
The slight difference in the amplitude of the meridional flow, due to 
the theoretical profile, explains the longer magnetic cycle of 40~yr
compared to 30~yr of \cite{Jouve+10b}.
The solution is antisymmetric and oscillatory, with concentrated strong polar regions,
and rather weak magnetic fields at low latitudes which are about $10^{-3}$ of the high-latitude fields,
migrating equatorward.
The model is labelled as ADV in Tab.~\ref{tab:setups} and falls short of
producing strong enough low-latitude toroidal fields for a realistic
solar butterfly diagram. Another issue is the cycle which remains too long.

This solution is located in the advection dominated regime, where the magnetic cycle
depends mostly on the meridional circulation.
The meridional circulation amplitude is defined as a byproduct of the
$\Lambda$-effect reproducing the solar differential rotation requiring only the mixing length
parameter $\alpha_{\rm MLT}$,
so the only possibility to decrease the magnetic cycle in this setup is to increase the turbulent
magnetic diffusion toward the diffusion dominated regime.

It is well known that dynamo solutions in the advection dominated regime differ from
the ones in the diffusion dominated regime. On the lower panel of Fig.~\ref{fig:non-delayed-model},
we show a model in the diffusion dominated regime, with $\eta_{\rm t} = 10^{12} {\rm cm}^{2} / {\rm s}$.
The relevant parameters can be found in Tab.~\ref{tab:setups} under the label DIFF.
The solution remains antisymmetric and oscillatory, with an activity cycle period of 8~yr.
The polar regions are less concentrated, closer to the solar characteristics and
the low latitudes exhibit stronger fields ($\approx~10$\% of the high-latitude fields), but the dynamo wave propagates
purely radially, as predicted by the Parker-Yoshimura rule.

Even if an $\eta_{\rm t}$ exists for which the non-delayed model gives a solution
with an 11-year cycle period, the low latitude radial fields remain too weak as compared to
the polar regions and the migration becomes radial while moving to the diffusion
dominated regime.

\section{Recovering the solar characteristics with the delay model}
\label{sec:the-delayed-dynamos}

We model the rise time as a temporal non-locality, the delay, because
\cite{Fournier+17} have shown in global simulations of rising magnetic
flux tubes that the rise time is independent of the turbulent magnetic diffusion.
In paragraph~\ref{sec:behavior-of-the-delay} we have shown that a time-dependent delay leads to temporal peaks in the source term
$S$ of the induction equation
near the surface, migrating latitudinally in the
direction of decreasing toroidal field.
We show below that the solar characteristics can only be recovered in
the diffusive regime (model D-DIFF).
We present one series of simulations in the advection dominated regime (D-ADV)
and another one in the 
diffusion dominated regime (D-DIFF),
In both series, $C_{\!S}$ and $\tau_0$ are varied.
Since we found that the accumulation is largest in the
nonlinear regime, we present two series with $\alpha = -2$.
The parameters ${\rm Re}$ and $C_\Omega$ are both fixed by the differential rotation
profile, whose shape is determined by the $\Lambda$-effect and a standard solar model.
$B_{\rm quench}$ and $B_{\rm threshold}$ are both constrained by the
results of \cite{Fournier+17} and $F_{\rm amp}$ by \cite{Hotta+12}.
The two remaining parameters are $C_{\!S}$ and $\tau_0$. Both
are not yet constrained by any global simulation results.
We have seen that $\tau_0$ may determine the variability of $S$ and $C_{\!S}$
its amplitude.
The parameters are summarized in Tab.~\ref{tab:setups}.

\subsection{The advection dominated case}

\begin{figure*}[tbph]
  \centering
    \includegraphics[width=1.0\linewidth]{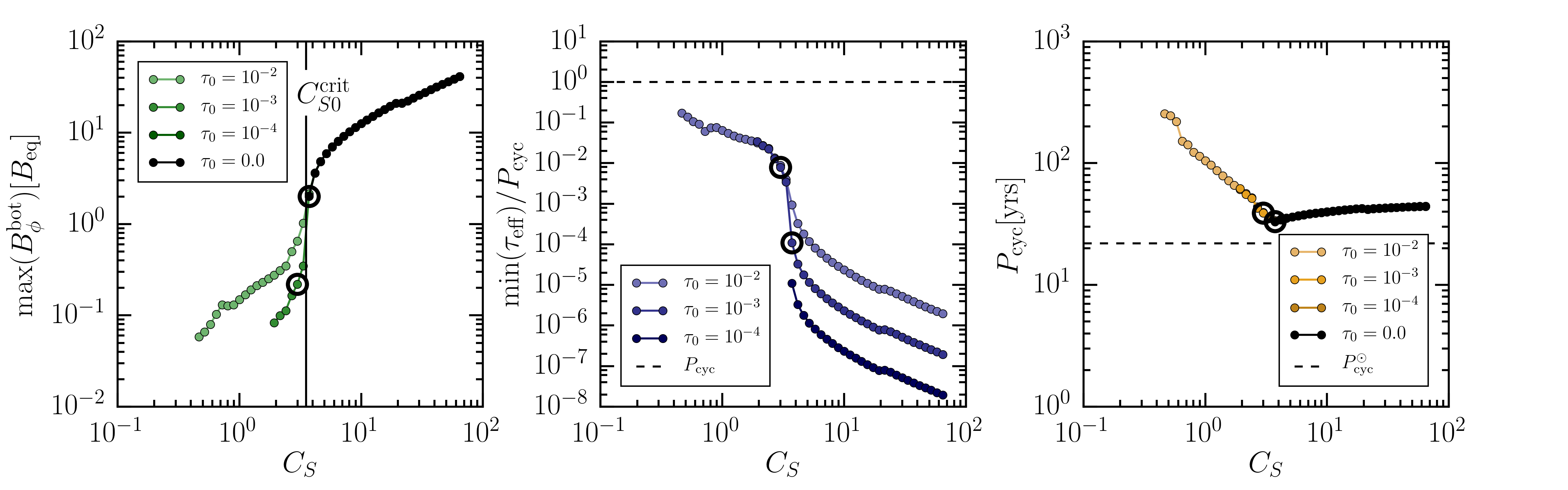}    
  \caption{
             \label{fig:strongly-non-linear-delayed-eta11-multi}
             D-ADV case. Saturation field strength (left), the effective delay (middle),
             and the cycle period (right), as functions of the source term amplitude $C_{\!S}$ for four different delays.
             Each line represents a series of runs for a given delay.
             The vertical solid line marks the critical $C_{\!S0}^{\rm crit}$ for the non-delayed case.
             The circles identify the solutions shown in \Fig{fig:strongly-non-linear-delayed-eta11-cont-br-surf-multi}.             
             }
\end{figure*}

In this section we consider the ADV setup, in the advection dominated regime, with $\eta_{\rm t} = 10^{11} {\rm cm}/{\rm s}$.
On the left panel of Fig.~\ref{fig:strongly-non-linear-delayed-eta11-multi}, we illustrate the maximum of the
toroidal magnetic field at the bottom of the convection zone, $B^{\rm cz}_{\phi}$, against $C_{\!S}$.
The non-delayed series with $\tau_0 = 0$ shows a critical source term amplitude of
$C^{\rm crit}_{\!S0} = 3.534$ below which the model produces decaying solutions.
It is remarkable that delayed dynamos deliver non-decaying solutions for weaker $C_{\!S}$ than $C^{\rm crit}_{\!S0}$.
The delay has the surprising effect of reducing the criticality of the dynamo,
opening a window to unknown solutions.

These solutions have the particular property to self-saturate at low amplitudes. The non-linearity
of the delay acts as a quenching mechanism, which is appealing since the solutions become independent
of the model used for the quenching. 

The middle panel of Fig.~\ref{fig:strongly-non-linear-delayed-eta11-multi} illustrates the dependence
of the relative effective delay ($\tau_{\rm eff} / P_{\rm cyc}$) on $\tau_0$ and $C_{\!S}$.
From this figure we could identify two different regimes:
the short-$\tau_{\rm eff}$ and the long-$\tau_{\rm eff}$ regime.

\begin{figure}[tbph]
  \centering
    \includegraphics[width=1.0\linewidth]{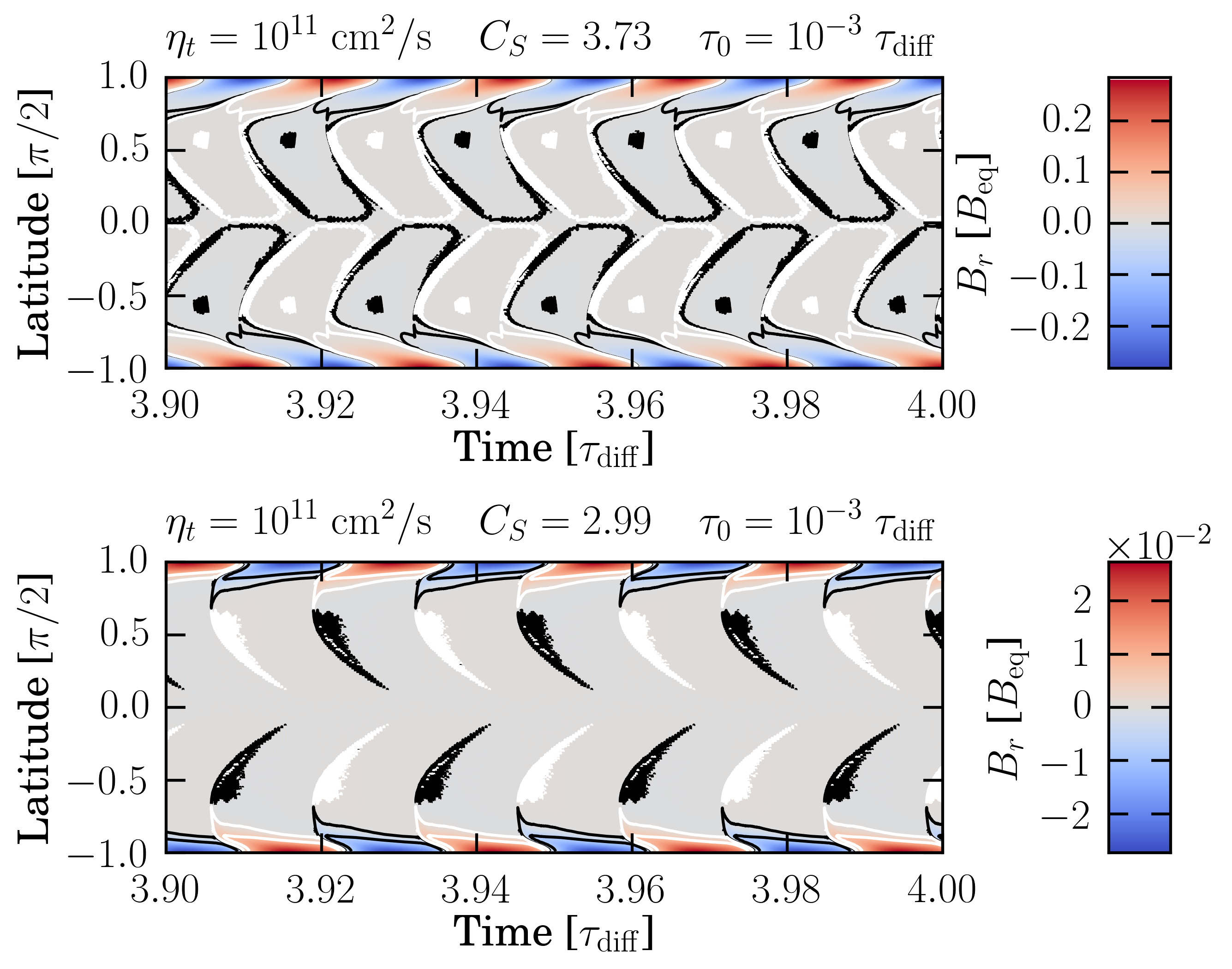}
    \caption{
             \label{fig:strongly-non-linear-delayed-eta11-cont-br-surf-multi}
             Radial magnetic field at the surface as a function of time from the advection dominated 
             case with delay (D-ADV) for two source term amplitudes $C_{\!S}$. 
             The respective parameters are indicated above the plot.
             The black solid lines represent the 
             1.5 and 0.15\% levels of $B_{r}^{\rm max}$ for the short-$\tau_{\rm eff}$
             case (top), and the 25, 2.5\% levels for the long-$\tau_{\rm eff}$ case
             (bottom).
             }
\end{figure}

In the short $\tau_{\rm eff}$ regime, the effective delay is several orders of magnitude shorter than the cycle period,
and therefore produces almost identical solutions to the non-delayed case.
Some accumulation is visible at mid-cycle (compare top panels of Fig.~\ref{fig:non-delayed-model}
and Fig.~\ref{fig:strongly-non-linear-delayed-eta11-cont-br-surf-multi}), for strongly non-linear delays ($\alpha = -2$) but
disappears for weaker non-linearity with $\alpha=-1$.
The cycle period is independent of $C_{\!S}$ as also shown in Fig.~\ref{fig:strongly-non-linear-delayed-eta11-multi}.

The second regime we identify is the long-$\tau_{\rm eff}$ regime which is characterized by $\tau_{\rm eff}$
between $10^{-3} P_{\rm cyc}$ and $10^{-1} P_{\rm cyc}$. 
This domain corresponds to the delay expected in the solar case -- between a few days and a few months -- and is
therefore the regime of interest. The solutions self-saturate leading to a strong dependence of the saturation
field and of the cycle period on $C_{\!S}$. The cycle period is also shown to be an increasing function of $\tau_{\rm eff}$.

It appears that the transient peaks in $S$ at the surface provide a sufficient source to maintain a dynamo
which would otherwise decay. Therefore the saturation mechanism is not the quenching of the source
term but the balance between the diffusion and the regular peaks of accumulated source term $S$.
We still do not understand why an increase of the effective delay (lower $C_{\!S}$) increases the cycle period.

\cite{Jouve+10b} studied the effect of $\tau_0$ for a given $C_{\!S} = C^{\rm crit}_{\!S0}$.
Here we demonstrate that the delay reduces $C^{\rm crit}_{\!S}$ and identify two regimes.
We studied the effect of $\tau_0$ for each regime and extended the analysis of \cite{Jouve+10b}.

In Fig.~\ref{fig:strongly-non-linear-delayed-eta11-multi}, it is remarkable that
the effective delay and the cycle period are independent of $\tau_0$. The
long-$\tau_{\rm eff}$ regime results from the non-linearity of the delay,
controlled by $\alpha$.
\cite{Fournier+17} showed that $\alpha$ is a function
of the azimuthal mode with which an unstable flux tube rises.
Determining under which condition one or the other mode is preferred
will provide solid input to constrain this parameter.

Fig.~\ref{fig:strongly-non-linear-delayed-eta11-cont-br-surf-multi} illustrates the
radial field at the surface in the long- and short-$\tau_{\rm eff}$ regimes.
In both regimes the morphological characteristics of the delayed dynamo resemble the non-delay case:
at high latitudes the strong polar fields are concentrated close to the pole propagating poleward;
at low latitude, the radial fields remain weak showing an equatorward propagation. 
Even though the accumulation in $S$ increases the field strength at
low latitudes, it remains two orders of magnitude weaker than the polar fields.

Additionally in the short-$\tau_{\rm eff}$ regime, the cycle period remains exclusively controlled
by the meridional circulation, like in the non-delayed solutions, but in the long-$\tau_{\rm eff}$ regime,
the cycle period clearly increases, with decreasing $C_{\!S}$.

We conclude that the non-linear delay does not affect the qualitative characteristics of
the dynamo, but, by increasing the cycle period, renders it a worse quantitative result.
We summarize the dynamo characteristics in table~\ref{tab:strongly-non-linear-delayed-summary}.

\subsection{The diffusion dominated case}

Estimates of the turbulent magnetic diffusivity from observations and 
mixing length based stellar models suggest a value of larger than $10^{12} {\rm cm}^2/{\rm s}$.
However as seen on the lower panel of Fig.~\ref{fig:non-delayed-model} for the non-delayed case in the diffusive regime,
even though the cycle period of about 8~yr fits the observations better, the low latitudes remain weakly active
and the dynamo wave propagates radially.
The fact that turbulent magnetic diffusivities of less than $10^{12} {\rm cm}^2/{\rm s}$ are required in 
order to obtain an equatorward migration has been a long-standing issue in Babcock-Leighton dynamos.

In Section~\ref{sec:behavior-of-the-delay}, we have shown that for a prescribed sinusoidal magnetic field,
the maximum field propagates in the latitudinal direction of the magnetic gradient, here we will
discuss this mechanism for the simulated results.

\begin{figure*}[tbph]
  \centering
    \includegraphics[width=1.0\linewidth]{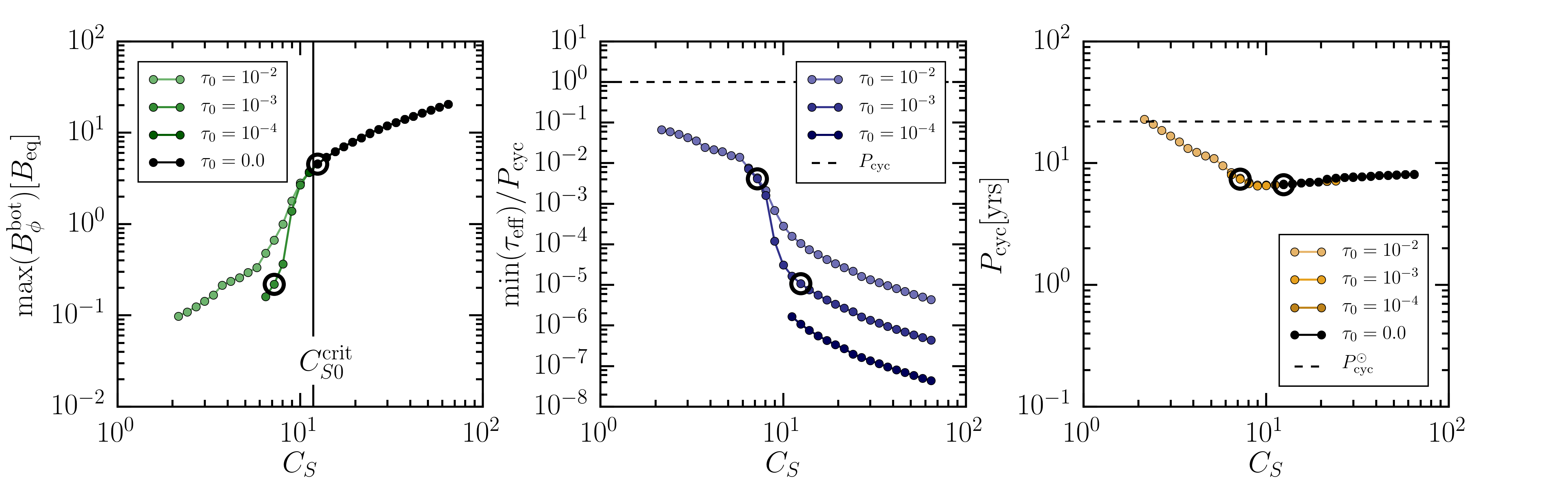}    
  \caption{
             \label{fig:strongly-non-linear-delayed-eta12-multi}
             D-DIFF case. Saturation field strength (left), the effective delay (middle),
             and the cycle period (right), as functions of the source term amplitude $C_{\!S}$ for four different delays.
             Each line represents a series of runs for a given delay.
             The vertical solid line marks the critical $C_{\!S0}^{\rm crit}$ for the non-delayed case.
             The circles identify solutions shown in \Fig{fig:strongly-non-linear-delayed-eta12-cont-br-surf-multi}.             
             }
\end{figure*}

\begin{figure}[tbph]
  \centering
    \includegraphics[width=1.0\linewidth]{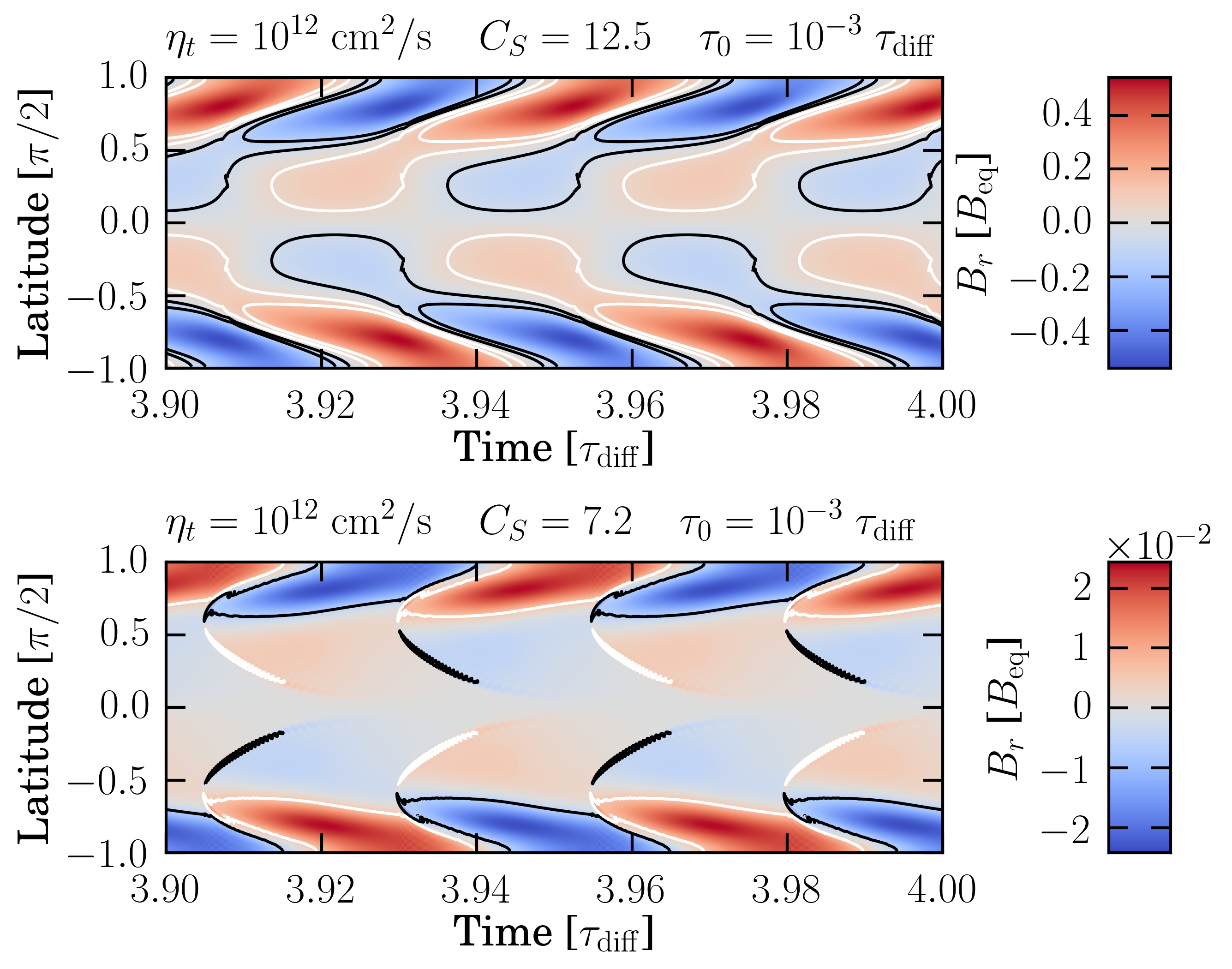}
    \caption{
             \label{fig:strongly-non-linear-delayed-eta12-cont-br-surf-multi}
             Radial magnetic field at the surface as a function of time from from the diffusion dominated 
             case with delay (D-DIFF) for $C_{\!S} > C_{\!S0}^{\rm crit}$ (top) and
             $C_{\!S} < C_{\!S0}^{\rm crit}$ (bottom).              
             The respective parameters are indicated above the plot.
             Contours represent the 25 and 10\% levels
             of $B_{r}^{\rm max}$ in the short-$\tau_{\rm eff}$ case (top),
             and the 25\% level in the long-$\tau_{\rm eff}$ case (bottom).
             }
\end{figure}

As in the advection dominated regime, the introduction of the delay reduces the criticality of the dynamo, and opens
a window towards new types of solutions. We could also identify the same regimes, illustrated in Fig.~\ref{fig:strongly-non-linear-delayed-eta12-multi}.
The short-$\tau_{\rm eff}$ regime remains identical to the non-delayed case, and
suffers from the same differences to the solar cycle (see upper panel of Fig.~\ref{fig:strongly-non-linear-delayed-eta12-cont-br-surf-multi}).
In the long-$\tau_{\rm eff}$ regime the overall behavior is similar, the cycle period and the effective delay are
independent of $\tau_0$. The solutions saturate before reaching $B_{\rm quench}$
(see left panel), and the cycle period decreases with $C_{\!S}$ (see right panel).
However, the morphologic characteristics of the dynamo are clearly different.
On the lower panels of Fig.~\ref{fig:strongly-non-linear-delayed-eta12-cont-br-surf-multi}, we can see how the reduction
of criticality allows an accumulation of $B_{r}$ reaching $\approx 25\%$ of the polar region's field strength.
Furthermore, the latitudinal distribution of the initial magnetic field (decreasing toward the equator) leads to an
equatorward propagation in the low latitudes. The last panel of Fig.~\ref{fig:strongly-non-linear-delayed-eta12-cont-br-surf-multi}
shows a solution which quantitatively agrees with the solar characteristics (see table~\ref{tab:strongly-non-linear-delayed-summary}).

In this particular run $\tau_{\rm eff} = 3~10^{-3} \, P_{\rm cyc}$ which is about a few days. But solutions
with effective delays up to a few months show comparable behaviors. So the non-locality introduces an additional time scale
of the order of days which is sufficient to obtain accumulation of radial-flux generation at low latitudes and equatorward migration.

\subsection{Comparison to the solar case}

\begin{table*}[tbh]
   \caption{ \label{tab:strongly-non-linear-delayed-summary} List of the characteristics of the dynamo, we have considered in this study.}
   \centering
\begin{tabular}{lccccc}
  \hline \hline
  ~                                         & Obs./Est.           & SOLAR               & D-ADV        & D-DIFF    & \\
  \hline 
  $\eta_{\rm t}$ [${\rm cm}^{2} / {\rm s}$] & $10^{10}$--$10^{14}$& $6.7 \cdot 10^{11}$ & $10^{11}$    & $10^{12}$ & \\
  $P_{\rm cyc}$ [${\rm yr}$]                & 11 (8--14)          & $\approx 11$        & 30--300      & 6--20     & \\
  low-latitude migration direction          & equatorward         & equatorward         & equatorward  & equatorward&\\
  high-latitude migration direction         & poleward            & poleward            & poleward     & poleward  & \\
  polar-field cap extent                    & $35\degr$           & $36\degr$           & $9\degr$     & $45\degr$ & \\
  low latitudes                             & active              & active              & low activity & active    & \\
  \hline 
\end{tabular}
\end{table*}

\begin{figure}[tbph!]
  \centering
  \includegraphics[width=1.0\linewidth]{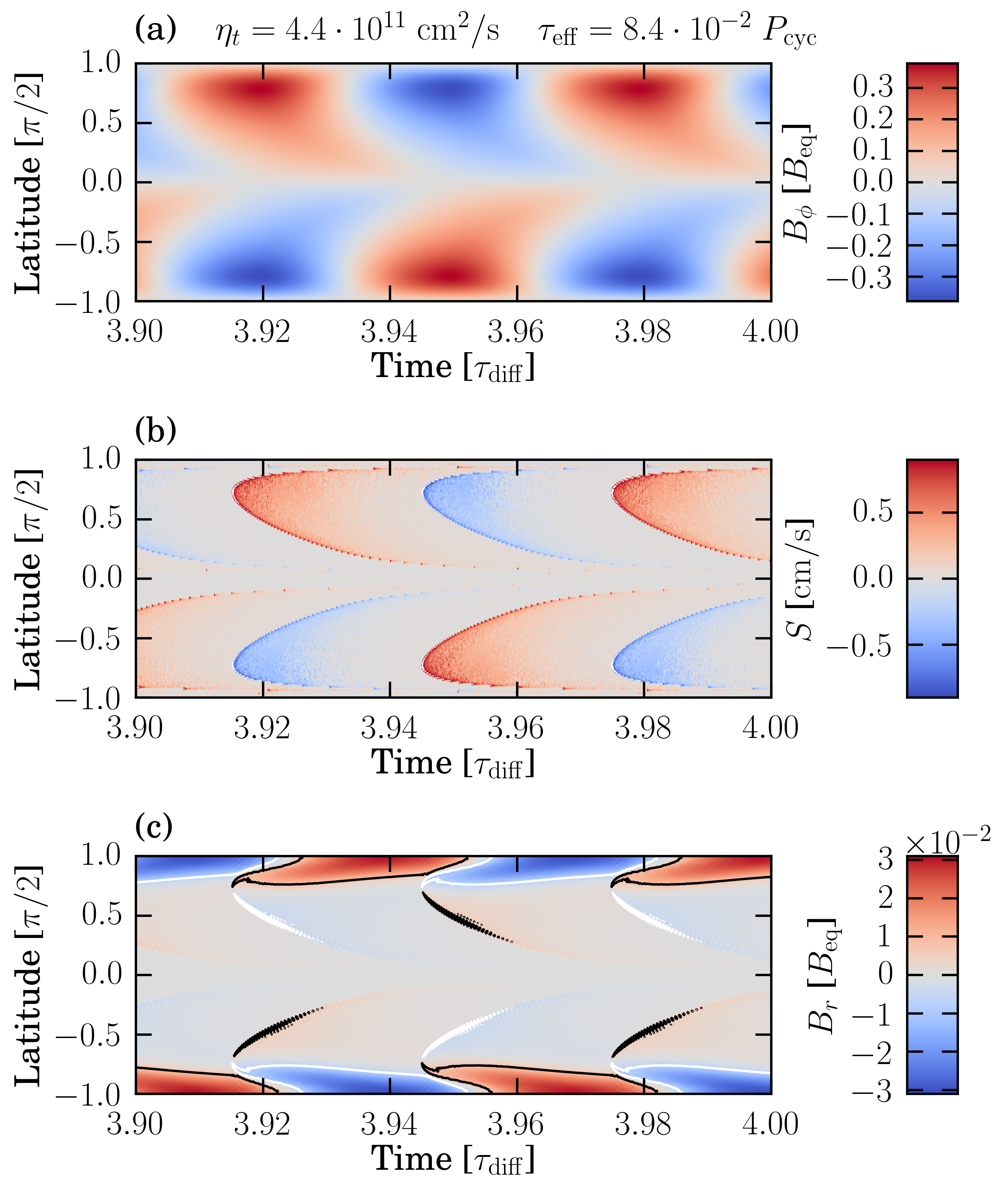}
\caption[]{
           \label{fig:solar-case-multi}
           Results of the SOLAR setup, with colour-coded 
           toroidal magnetic field at the bottom of the convection zone (a), the
           resulting delayed source term at the surface (b) and the radial magnetic
           field at the surface (c). Contours in panel
           (c) represent the 25\% level of $B_{r}^{\rm max}$.
          }
\end{figure}

The series we carried out in the diffusion dominated regime
shows already a good quantitative agreement with the solar case, but we scanned the constrained parameter space
with the remaining free parameters, namely $\eta_{\rm t}$, $\tau_0$ and $C_{\!S}$, and selected
a simulation, referred to as SOLAR, which reproduces the solar characteristics.
Varying $C_{\!S}$ and $\eta_{\rm t}$, we could find a dynamo solutions whose butterfly diagram
matches quantitatively several characteritics of the solar observations. We summarized these aspects
in Tab.~\ref{tab:strongly-non-linear-delayed-summary}.
We have been comparing, the activity cycle period, the propagation of the active belt and of the high-latitude,
the extent of the polar regions as well as the activity level of the low latitudes.
We would like to remind the reader that the averaged strength of the observed active belts is about an order of magnitude
weaker than the polar field strength.

The SOLAR solution is antisymmetric and oscillatory, with a cycle of 11~yr, $36\degr$ extent of
the polar regions, with the amplitude of low latitudes being a fourth of the amplitudes of the polar regions, 
and an equatorward propagation of the active belt as well as a poleward propagation of the
high-latitude fields. It is also remarkable that the polar reversal happens at half-cycle of the low latitude.
The turbulent magnetic diffusivity required to obtain such a solution is $\eta_{\rm t} = 6.7\cdot10^{11}~{\rm cm}^{2}/{\rm s}$ which
is in the transitional regime between the diffusive and the advective regime. The
resulting effective delay $\tau_{\rm eff}$ is of the order of a month ($\approx 33$~days).
The toroidal field at the bottom of the convection zone saturates below equipartition with the
convective motions, at about $0.15 B_{\rm eq}$.

\begin{figure}[tbph!]
  \centering
  \begin{tabular}{ll}
    \includegraphics[width=0.4\linewidth]{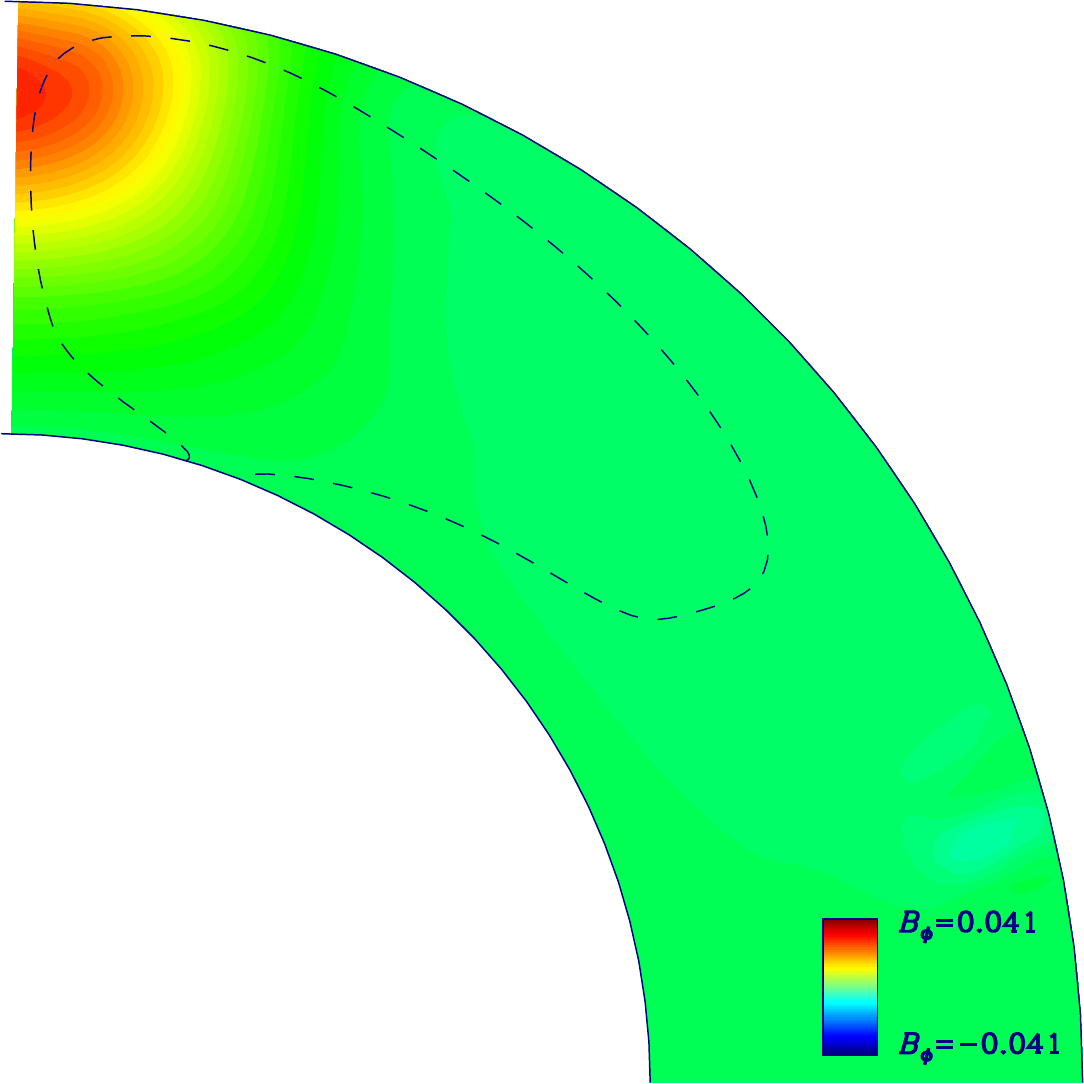} & \includegraphics[width=0.4\linewidth]{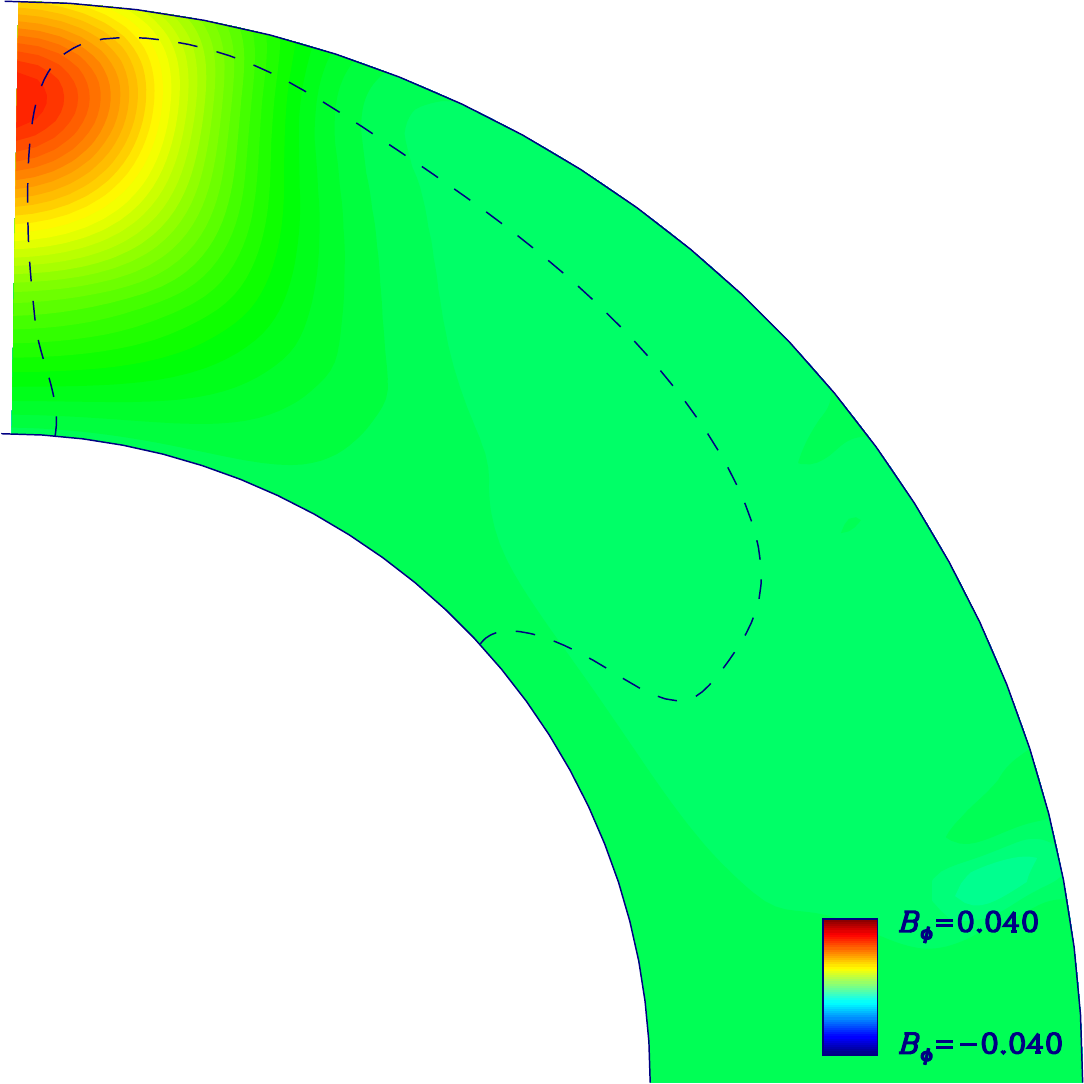} \\
    \includegraphics[width=0.4\linewidth]{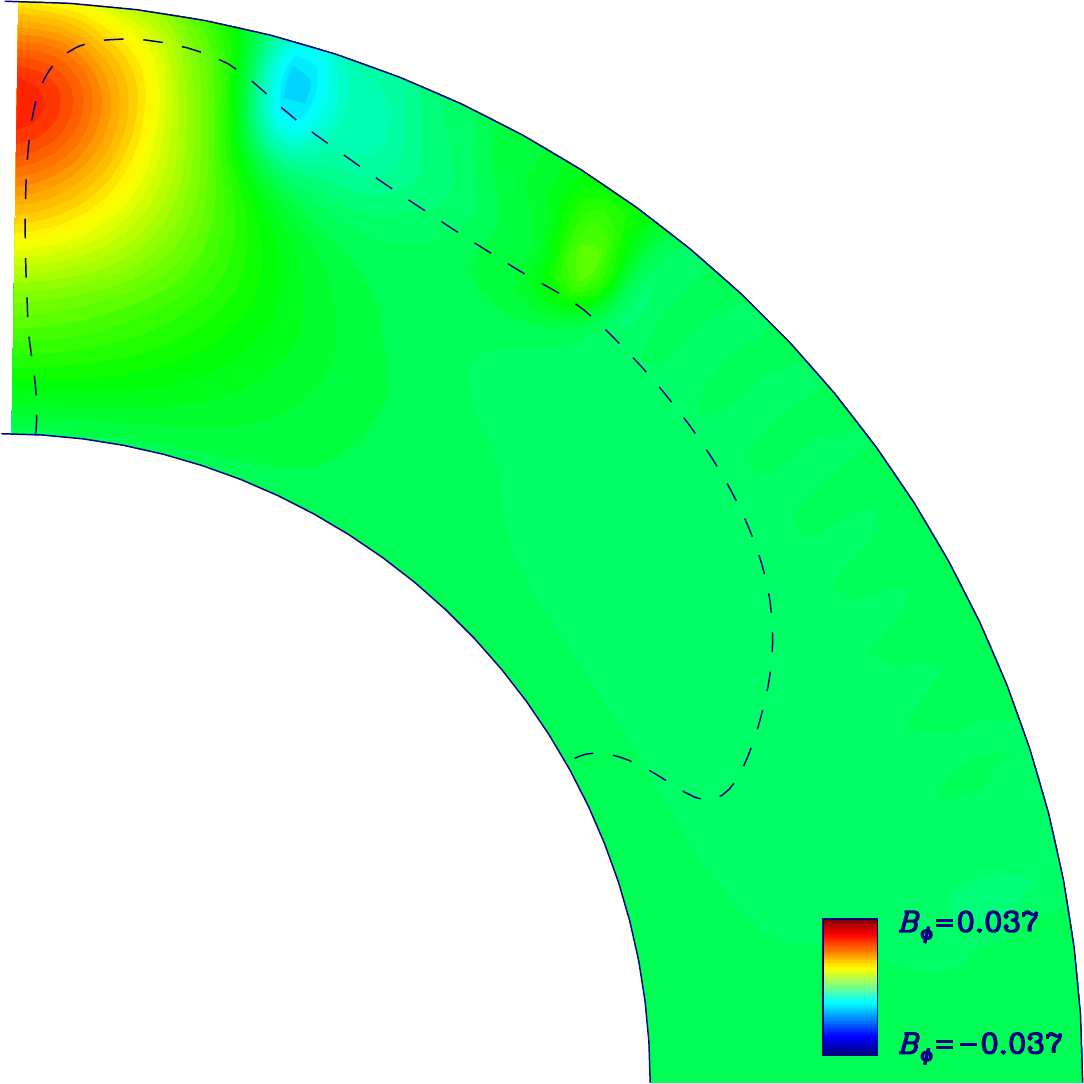} & \includegraphics[width=0.4\linewidth]{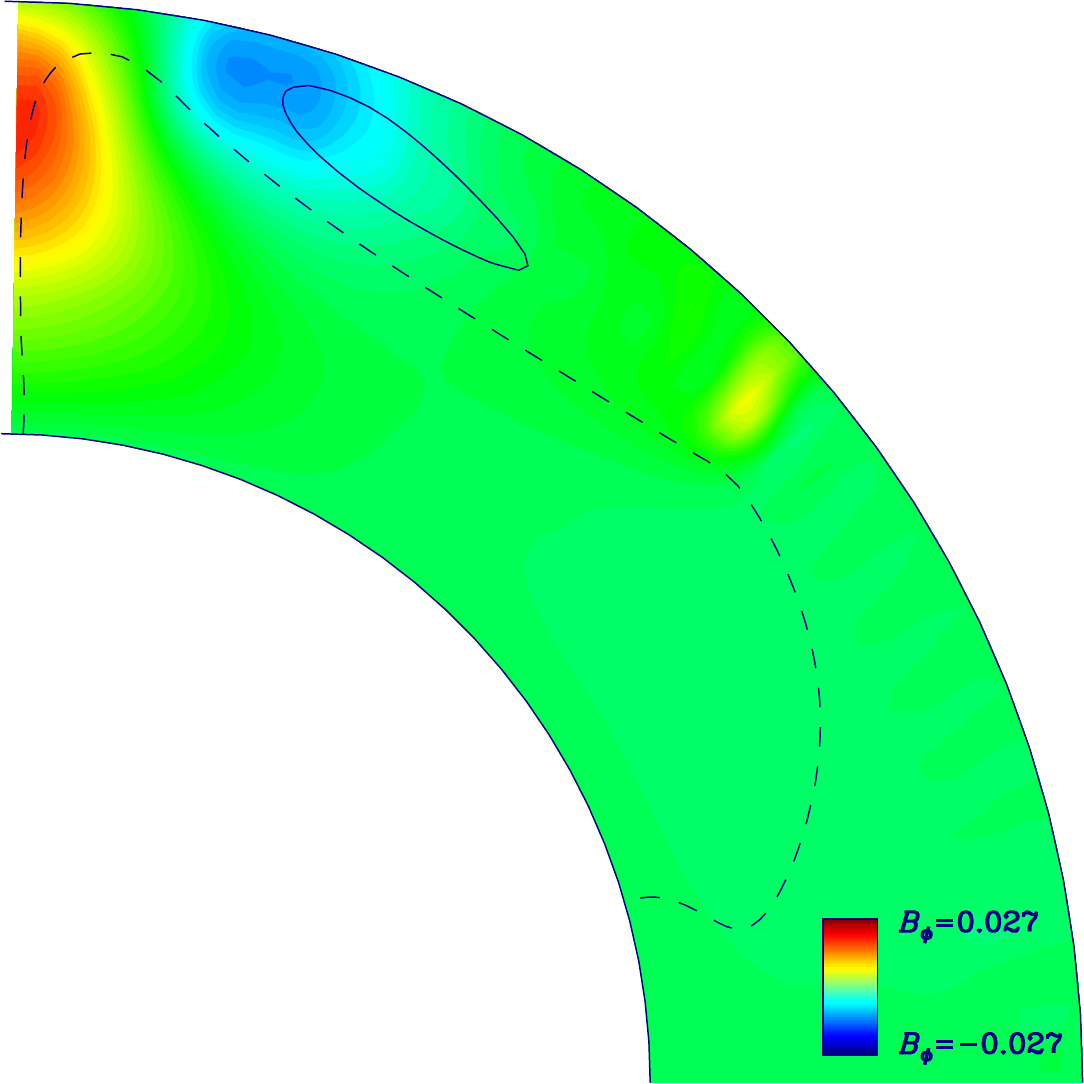} \\
    \includegraphics[width=0.4\linewidth]{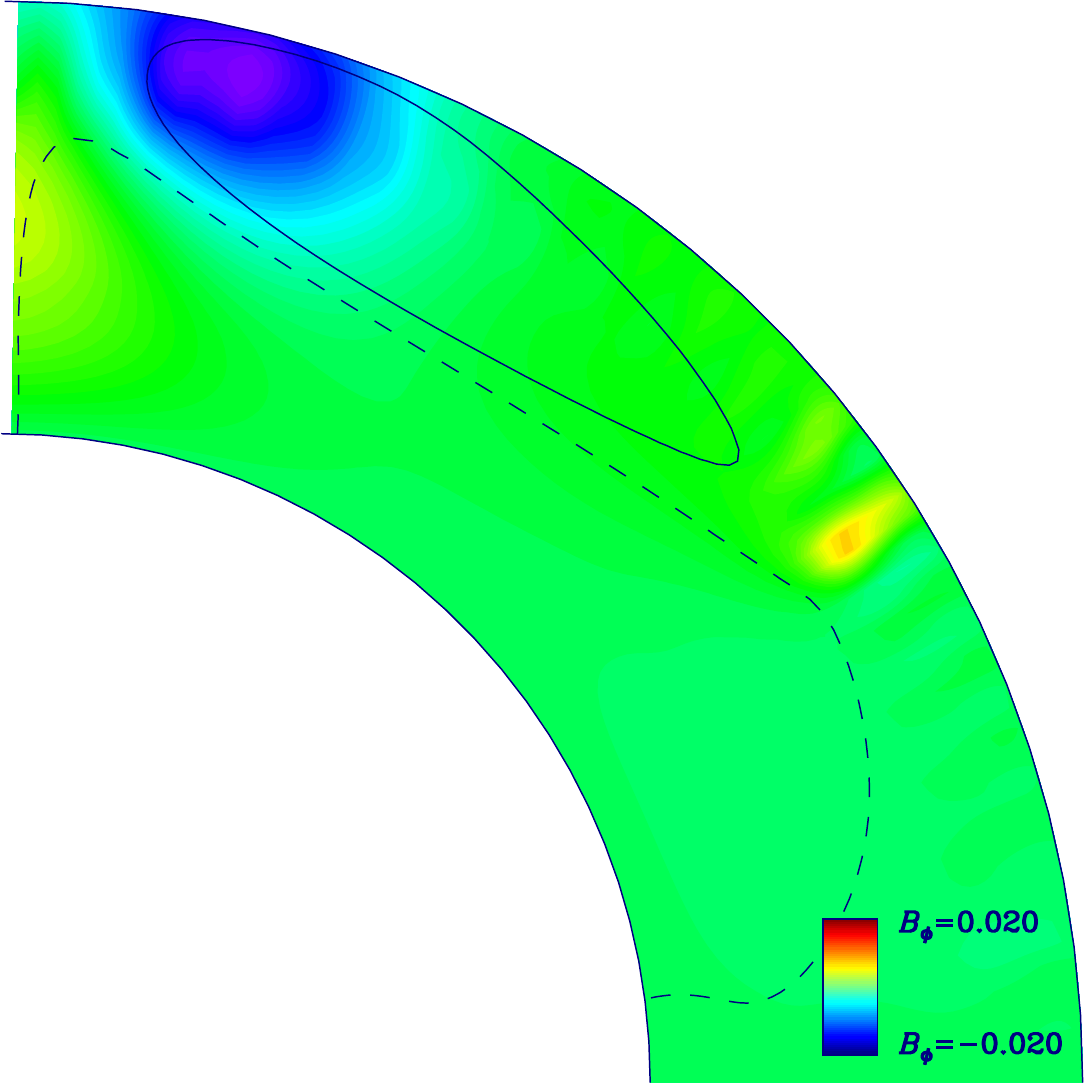} & \includegraphics[width=0.4\linewidth]{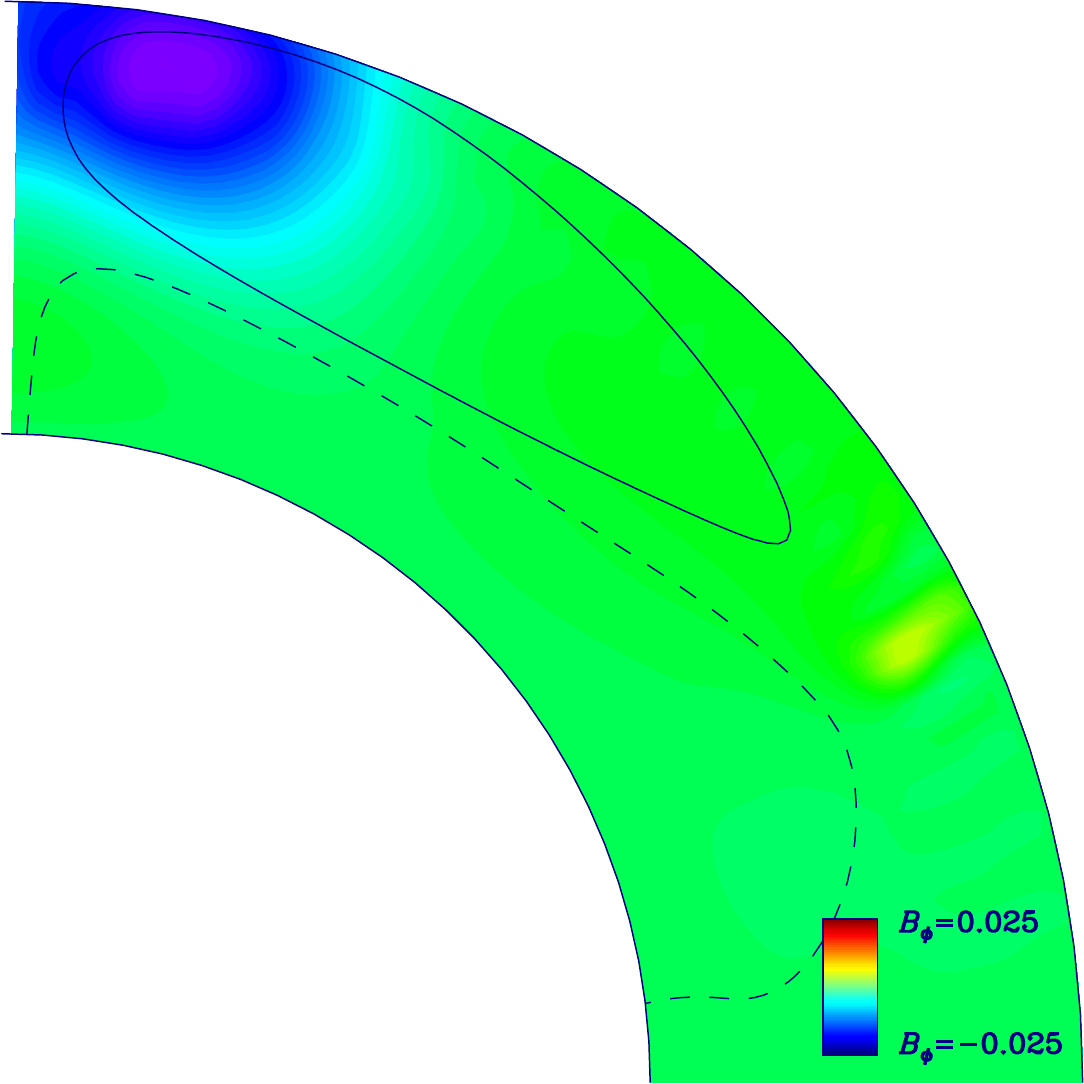} \\
    \includegraphics[width=0.4\linewidth]{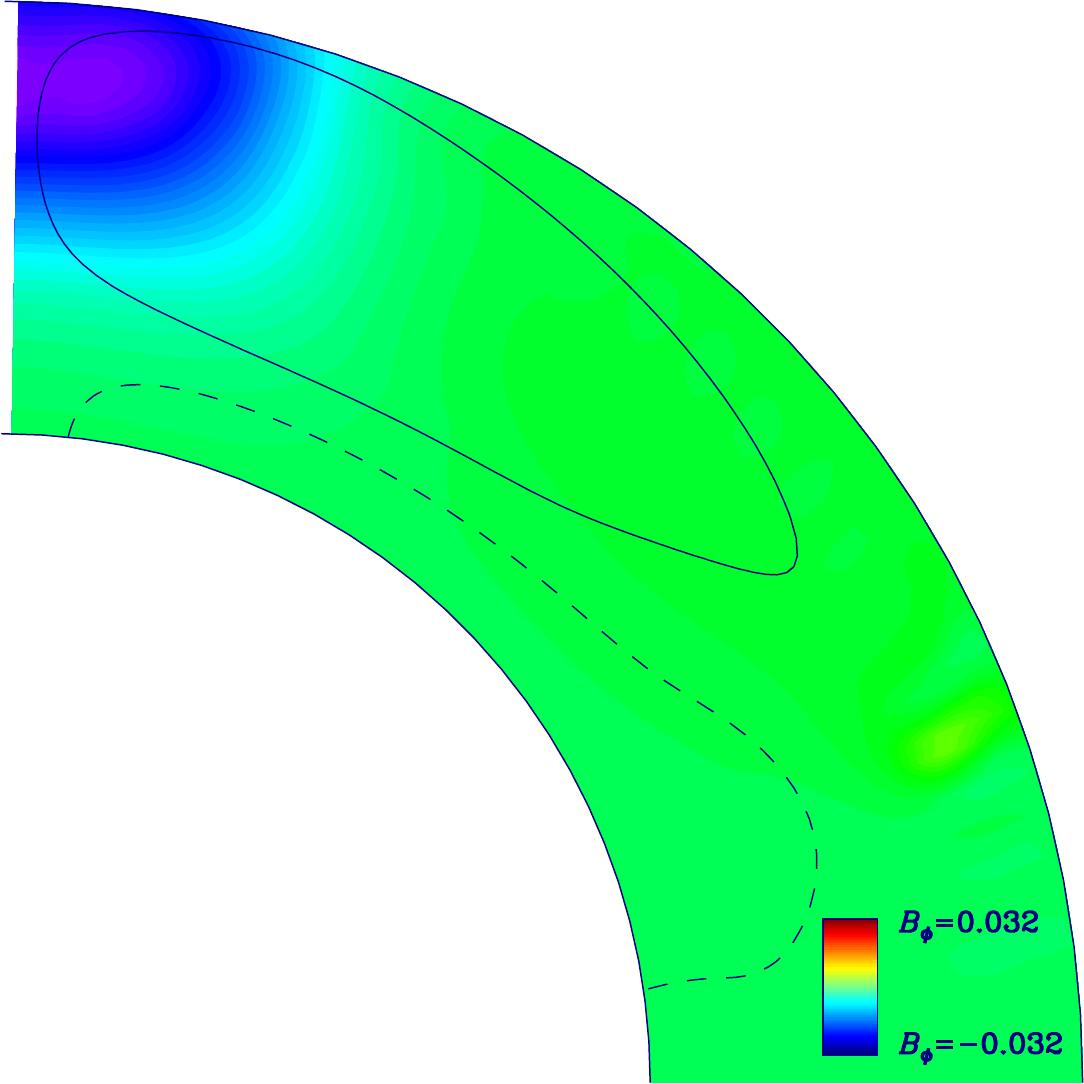} & \includegraphics[width=0.4\linewidth]{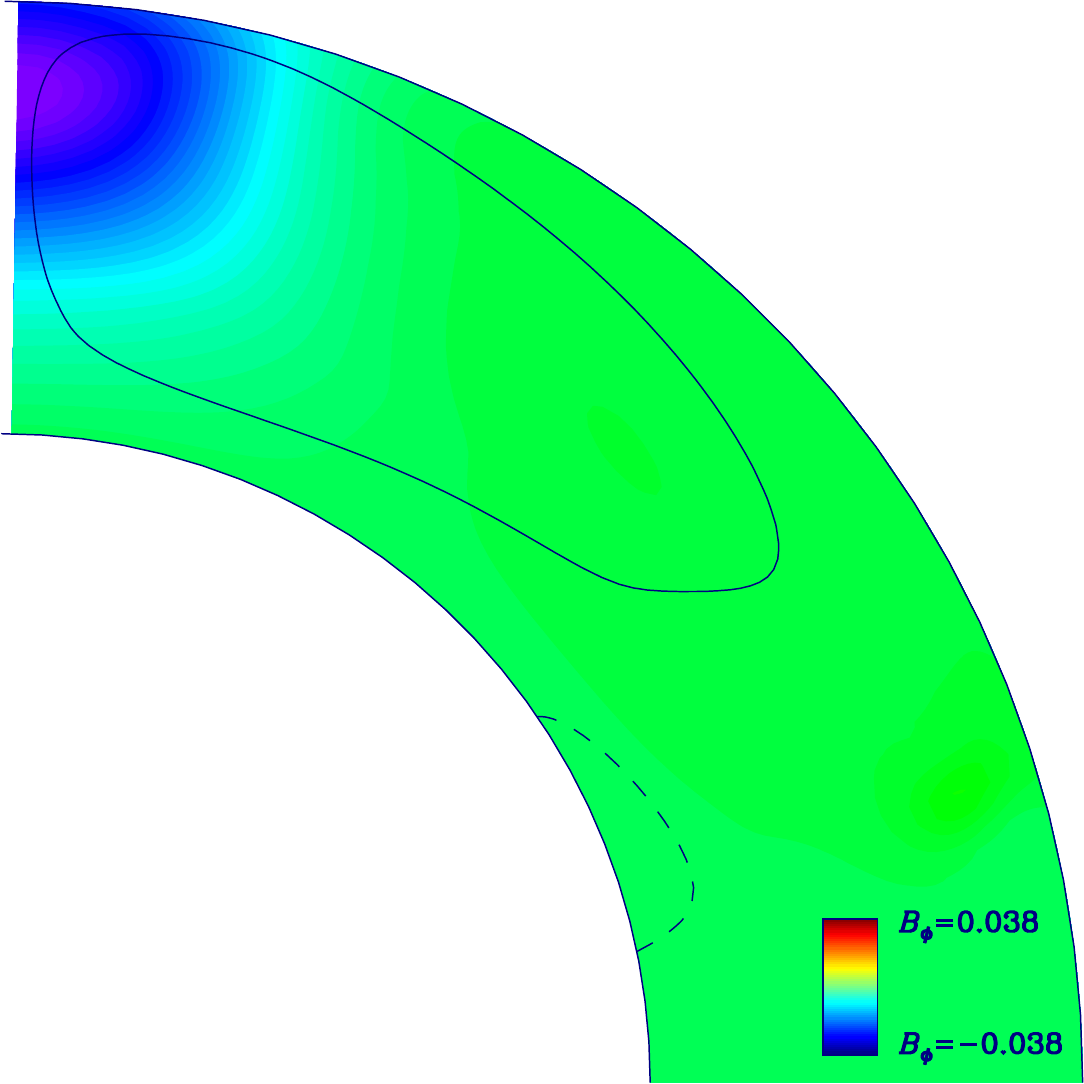} \\    
  \end{tabular}
\caption[]{
           \label{fig:solar-case-slices}
           Slices of the northern hemisphere. The slices are taken at different phases of the cycle,
           ordered from left to right and top to bottom: the rising phase, the maximum phase, the declining phase, and
           the minimum of the activity cycle. The colored contours represent the radial magnetic field, with
           red pointing outward and blue inward respectively.
           Contours are 5\% and 2.5\% of the maximum $B_{\phi}$.
          }
\end{figure}

In Fig.~\ref{fig:solar-case-multi}, we illustrate various aspects of the dynamo mechanism
showing how the toroidal field profile (a) is distorted
into a source term sharply peaked in time (b), building up a ``front''
which, added to the diffuse field at the surface, leads to the very characteristics of the butterfly diagram (c).
Because of the stiff accumulation, the active latitudes are strongly localized in latitude and time as compared to the solar butterfly diagram.
As \cite{Weber+11} have shown that convective motions introduce stochasticity in the emergence characteristics of active regions,
this mechanism could explain the broadening of the real solar activity bands in the butterfly diagram.
We illustrate the propagation of the active latitudes in the meridional plane in Fig.~\ref{fig:solar-case-slices} where the solid and dashed
black contours represent the toroidal field, and the colour-coding shows the negative and positive radial field strength, respectively.
The slices are taken in the beginning of the activity cycle, at maximum, in the decreasing phase and at minimum. 
The toroidal field peaks at high latitudes and propagates almost radially, as predicted
by the Parker-Yoshimura rule. The strongest radial field is located at the pole. Close to the surface,
where the source term is the strongest, the active latitudes, indicated by the dashed ovals, can be seen to migrate to the equator. 
These meridional sections demonstrate that the equatorward migration of the active belt does not
follow the migration of the dynamo wave
which propagates almost radially, but results from the longer delay for weaker toroidal fields closer to the equator.

\section{Discussion of robustness of the results}

Since some parameters remain weakly or even not constrained, it is
important to discuss the robustness of the results of the previous Section.

The accumulation of the source term $S$ in short-lived peaks due to the
delay is the key aspect of the model which provides the solar-like characteristics.
And for a prescribed oscillating magnetic field the accumulation is controlled
by two parameters $\tau_0$ and $\alpha$ (see Section~\ref{sec:behavior-of-the-delay}).
\cite{Fournier+17} could constrain $\alpha$ in the range of $-0.91$ to
$-2$ depending on the unstable azimuthal mode of the rising flux tube.
However one of the limitations of the latter work is the lack of
constraint on $\tau_0$. It depends on many details of the formation
of magnetic flux tubes and will be clearly challenging to constrain.
Fortunately, we could demonstrate that the actual solutions in the
long-$\tau_{\rm eff}$ regime are clearly independent of $\tau_0$.
As for the dependence on $\alpha$ we have varied
$\alpha$ from $-2$ to $-0.1$ and found that solar-like solutions
could be obtained until $\alpha = -1$, which remains
within the above constrained space. For $-1 < \alpha$ the accumulation is not
sufficient and the dynamo decays.

Even though we could constrain the value of $B_{\rm quench}$ in the first
term in the sum of (\ref{eq:source-term}), thanks to global
numerical simulations, the quadratic quenching we use is quite arbitrary.
But since this model possesses the surprising capability of self-quenching because of its non-linear
non-locality, the solutions are independent of the chosen quenching model.
The level of saturation only depends on the effective delay, $\tau_{\rm eff}$, which is
an outcome of the model and cannot be chosen arbitrarily.

Because we use a lower threshold on the magnetic field, some mode may not grow.
The preferred mode of the dynamo is determined by the choice of the threshold and the initial condition.
Although the threshold is relatively well constrained from the stability analysis of the buoyancy
instability \citep{Ferriz-mas+94},
it prevents the dynamo to grow from an arbitrary low seed field.

As is expected for the diffusion dominated regime, the cycle period depends on
$\eta_{\rm t}$, but in contrast to the non-delayed dynamos, the delayed model additionally
shows a dependence on $C_{\!S}$. The cycle period may change by a factor of two over the relevant range of $C_{\!S}$.
This is remarkable because it could explain how solar-like stars with a comparable
rotation period and convective envelope (same $\eta_{\rm t}$) could show
different magnetic cycles -- due to a difference in metalicity for instance.
Such dependence needs to be carefully addressed because
the interpretation of $C_{\!S}$ as a physical quantity is not trivial.
More global simulations will be required to be conclusive on this issue. But
the non-local dynamos seems to be good candidates to address this particular issue.

\section{Conclusions}
\label{sec:discussion-and-conclusions}

Until now, diffusive Babcock-Leighton dynamos were considered
not to be able to reproduce qualitatively the solar dynamo. The
Parker-Yoshimura rule implies for the internal differential rotation
of the Sun that the dynamo wave propagates poleward.
We also find The cycle period to be too short as well as the low latitude
radial fields to be too weak.

In the present work we introduced a delay in the source term of the
poloidal field. Like in \cite{Jouve+10b}, this delay represents
the rise time of magnetic flux tubes through the convection zone.
But in contrast to former studies, we built this model on the results of global
numerical simulations of rising magnetic flux tubes in compressible stellar
interiors \citep{Fournier+17}. The model consists of a rise time which
depends nonlinearly on the magnetic flux density.

We have shown that the nonlinearity of the delay leads to an
accumulation of the Babcock-Leighton source term at certain times. When
this accumulation becomes sufficiently important, it may prevent
the dynamo from decaying, even though the non-delayed model shows no dynamo action. The reduction of the criticality of
the dynamo opens a new window to unknown solutions.
These delayed dynamos have the peculiar property of self-quenching.

We found that the nonlinear delay can provide a mechanism to
generate migration of the surface fields in the direction of weaker internal fields. 
In case of a stationary toroidal internal field at mid-latitudes, for example, the 
generated poloidal fields at the surface migrate towards the equator at low latitudes
and towards the poles at high latitudes. This is independent of the sign of the
internal differential rotation.

The requirement of a low turbulent magnetic diffusivity, $\eta_{\rm t}$,
for Babcock-Leighton dynamos to reproduce qualitatively the solar cycle,
has been shown to be unnecessary. We demonstrate that the present delayed model,
with a turbulent magnetic diffusivity of $\eta_{\rm t} = 6.7\cdot10^{11} {\rm cm}^{2}/{\rm s}$,
agrees well with the solar butterfly diagram, even though the diffusivity
is relatively high throughout the entire convection zone. 

Note that one proposed way out of the low-diffusivity problem is to 
use different values for $\eta_{\rm t}$ for the toroidal and for the
poloidal components in the induction equation \citep{Chatterjee+04}.
While the diffusivity may well be different in the horizontal and
vertical directions, the poloidal field has varying components in both 
the horizontal and vertical directions, rendering the poloidal-field
diffusivity location-dependent. We have not tried such a setup, and think it is
actually not necessary given the results presented.

In any case, the model presented in this work is, by design, a simplified model.
It has allowed us to identify the effect of the delay on the dynamo
solutions. However several ingredients are missing to reach a state-of-the-art
model
\citep{Rempel06, Cameron+15, Pipin17}.
We only solve the induction equation for large-scale fields.
We ignore 
the turbulent pumping,
and the back-reaction of the magnetic field on the flow.
All these elements will increase the complexity of the model and bring along
additional free parameters which need to be constrained.

The large-scale field generation based on the Babcock-Leighton effect
has not been derived from first principles. Its validity remains therefore
uncertain.
In the absence of global simulations addressing the formation of
magnetic flux tubes,
the current models remain quite arbitrary.
Nevertheless, the non-linearities of the presented solutions
are potentially relevant for other dynamos than the Babcock-Leighton type.

Finally, the relevance of this work for stellar dynamos will
be revealed only if this model is proven to robustly reproduce observed
dynamo patterns of further solar-like stars.

\section{Acknowledgment}
\label{sec:acknowledgment}

We would like to thank the participants and organizers of the
Natural Dynamos (2016) conference for constructive remarks and input.

\bibliographystyle{aa}
\bibliography{references}

\appendix

\section{Constraining $B_{\rm quench}$ and $B_{\rm threshold}$}

The quenching field strength is defined such that the buoyant force balances
the Coriolis force, resulting into a zero tilt at the surface.
\citet[][Section 3]{Fournier+17} showed for the axisymmetric case that
\begin{equation}
  \frac{F_{\rm buoy}}{F_{\rm corio}} = 0.7 \, \Gamma_{1}^{1} 
\end{equation}
with
\begin{equation}
  \Gamma_{1}^{1} = \frac{v_{\rm A}}{\varpi \Omega}
\end{equation}
Presuming that the ratio of the buoyant force over the Coriolis force is unity
for the quenching field strength, one obtains:
\begin{equation}
  B_{\rm quench} = \frac{1}{0.7} \, \varpi \Omega \, \sqrt{ \rho \mu_0} = 32 \cdot 10^{4} {\rm G} = 32 \, B_{\rm eq}
\end{equation}
with $\Omega = 430~{\rm nHz}$, $\rho = 0.1 {\rm g}/{\rm cm}^3$, $\varpi = 0.7 ~ 2 \pi ~ 7 \cdot 10^{10} {\rm cm}$ and
$B_{\rm eq} = \sqrt{\mu_0 \rho u^{2}_{\rm rms}} = 10^{4} {\rm G}$ with $u_{\rm rms} = 100 {\rm m}/{\rm s}$.

The field strength of a magnetic flux tube which reaches the surface needs to be larger than $3 B_{\rm eq}$ \citep{Fan+03}.
The threshold field strength is therefore $3 B_{\rm eq}$.

\end{document}